\documentclass[twocolumn,oldversion]{aa}
\usepackage{graphicx}
\usepackage{txfonts}
\usepackage{natbib}
\bibpunct{(}{)}{;}{a}{}{,}
\newcommand{\vcas}  {4U 0115+63}
\newcommand{\exo}   {EXO 2030+375}
\newcommand{\bq}    {V0332+53}
\newcommand{\ks}    {KS 1947+300}

\def\simless{\mathbin{\lower 3pt\hbox
     {$\rlap{\raise 5pt\hbox{$\char'074$}}\mathchar"7218$}}}   
\def\simmore{\mathbin{\lower 3pt\hbox
     {$\rlap{\raise 5pt\hbox{$\char'076$}}\mathchar"7218$}}}   

\begin{document}

   \title{Rapid spectral and timing variability of Be/X-ray binaries during type II 
   outbursts \\
   }

   \subtitle{}

   \author{P. Reig
          \inst{1,2}
          }

\authorrunning{Reig et~al.}
\titlerunning{Rapid variability of Be/X-ray binaries}

   \offprints{P. Reig}

   \institute{IESL, Foundation for Research and Technology-Hellas, 71110,
   Heraklion, Greece 
   \and
   	Physics Department, University of Crete, 71003, Heraklion, Greece
              \email{pau@physics.uoc.gr}
             }

   \date{Received ; accepted}

\abstract{
X-ray colour-colour (CD) and colour-intensity (HID) diagrams are powerful
tools that allow the investigation of spectral variability without the
assumption of any spectral model. These diagrams have been used extensively
in low-mass X-ray binaries and black-hole candidates but very few
applications are found for high-mass X-ray binaries. We have investigated
the spectral and timing variability of four accreting  X-ray pulsars with
Be-type companions during major X-ray outbursts.  The aim is to define 
source states based on the properties (noise components) of the aperiodic 
variability in correlation with the position in the colour-colour diagram. 
Different spectral states were defined according to the value of the X-ray
colours and flux. Transient Be/X-ray binaries exhibit two branches in their
colour-colour and colour-intensity diagrams: the horizontal branch
corresponds to a low-intensity state and shows the larger fractional $rms$,
similar to the the island state in atolls and horizontal branch in Z
sources;  the diagonal branch corresponds to a high-intensity state, in
which the source spends about 75\% of the total duration of the outburst.
Despite the complexity of the  power spectra due to the peaks of the pulse
period and its harmonics, the aperiodic variability of Be/X-ray  binaries
can be described with a relatively low number of Lorentzian components.
Some of these components can be associated with the same type of noise as
that seen in  low-mass X-ray binaries, although the characteristic
frequencies are about  one order of magnitude lower.   The analysis of
the CD/HID and power spectra results in two different types of Be/X.  While
in \vcas, \ks\ and \exo\ the hard colour  decreases as the count rate
decreases, in \bq\ it increases.  The pattern traced by \bq\ then results
in a Z shaped track, similar to the low-mass $Z$ sources, without the
flaring branch. In contrast, the horizontal branch in \vcas, \ks\ and \exo\
corresponds to a low/soft state, not seen in other types of X-ray binaries.
The noise at very low frequencies follows a power law in \bq\ (like in LMXB
Z) and it is flat-topped in \vcas, \ks\ and \exo\ (like in LMXB atoll). \bq\
shows a noise component coupled with the periodic variability that it is
not seen in any of the other three sources.

}


   \maketitle

\begin{table*}
\caption{Optical and X-ray information of the systems and the outbursts analysed 
in this work.}             
\label{info}      
\centering          
\begin{tabular}{lcccccccc}
\hline\hline
Source		&Spectral&P$_{\rm spin}$ &P$_{\rm orb}$ &$e$	&Distance &Outburst	&$L_{\rm X,peak}$ &On-source   \\
name		&type	&(s)   		&(days)		&	&(kpc)	&duration (d)  &3-30 keV (erg s$^{-1}$) &time (ks)    \\
\hline
4U 0115+63	&B0.2V	&3.6	&24.3	&0.34 	&8.1	&55	&$1.2\times10^{38}$	&91	\\
KS 1947+300	&B0V	&18.7	&40.4	&0.03 	&10	&165	&$6.2\times10^{37}$	&132	\\
EXO 2030+375	&B0.5III-V&41.8	&46.0	&0.41	&7.1	&155	&$1.4\times10^{38}$	&166	\\
V 0332+53	&O8-9V	&4.4	&34.2	&0.30 	&7	&105	&$3.2\times10^{38}$	&198	\\
\hline\hline
\end{tabular}
\end{table*}

\section{Introduction}

The colour-colour (CD) and hardness-intensity (HID) diagrams are very
useful tools to investigate the spectral variations of an X-ray source.
They are model- and instrument-independent and reflect the intrinsic
properties of the system. The first application of this type of plots was
the separation of different types of galactic X-ray sources and the
classification of X-ray populations.  The spectral hardness was defined by
the ratio of counting rates in each of the proportional counters of the
{\em MIT OSO-7} experiment \citep{mark77,rapp77}. \citet{whit84} made use
of the X-ray colour-colour diagrams, obtained from {\em HEAO 1}
observations, to study the spectral states among the different types of
X-ray binaries, including black-hole candidates.

CDs became widely used during the EXOSAT era, which also marks the 
introduction of HIDs \citep{pried86,klis87a,hasi87}, where "hardness"
refers to the ratio of the count rate in two energy bands. Before EXOSAT,
CDs contained sources of different origin. The novelty of the EXOSAT CDs
and HIDs was that they were applied to individual sources. CD/HIDs became a
useful way to investigate the rapid aperiodic variability in correlation
with spectral states, as defined by the spectral hardness, of individual
low-mass X-ray binaries and black-hole candidates \citep{schu89,hasi89}.

With the advent of RXTE the use of CDs and HIDs have become part of the
standard data analysis. RXTE has provided an unprecedentedly large database
of X-ray observations with exceptionally good timing and moderate spectral
resolution.  Together with the functional form of the variability
components ("noise"), CD and HID have become an essential tool in
developing a whole new phenomenology on the spectral and timing properties
of X-ray binaries by introducing the notion of source states.  A state is
defined by the appearance of a spectral (i.e. power-law, blackbody) or
variability component (i.e. Lorentzian) associated with a particular and
well-defined position of the source in the CD/HID. A transition between
states is assumed when the relative strength of the spectral or variability
components varies and the source motion in the CD/HID changes direction
\citep{klis06}.

While there are numerous references in the literature on the application of
CD/HID analysis on low-mass X-ray binaries \citep[][and references
therein]{klis06} and black-hole systems \citep[see e.g.][]{bell05},  very
little work of this type has been done on high-mass X-ray binaries.
\citet{bell90} performed an aperiodic variability study of 12 high-mass
X-ray binaries from EXOSAT archive data. The sample included \exo\ and \bq.
However, the few observations analysed of \bq\ did not correspond to a
major X-ray outburst, while the data of \exo\ covered the decay of the 1985
outburst only partially. The other 10 sources were supergiant X-ray
binaries in bright states. The vast majority of the rapid aperiodic
variability studies in accreting X-ray pulsars have concentrated on the
detection of QPOs \citep{ange89,jern00,qu05}; 
see also \citet[][and references therein]{shir02}

Massive X-ray binaries are classified according to the luminosity class of
the optical component into Be/X-ray binaries (dwarf or subgiant) and
supergiant (luminosity class I-II) X-ray binaries. The former tend to be
transient systems while the latter are persistent sources. Be stars
are non-supergiant fast-rotating B-type and luminosity class III-V stars
which at some point of their lives have shown spectral lines in emission.
In the infrared they are brighter than their non-emitting  counterparts of
the same spectral type. The line emission and infrared excess
originate in extended circumstellar envelopes of ionized gas surrounding
the equator of the B star. When the Be star takes part on a binary system
where the companion is a neutron star, then the system is referred to as a
Be/X-ray binary.

Be/X-ray binaries are variable on time scales from seconds to years. The
fastest variability is found in the X-ray band. Virtually all the Be/X-ray
binaries with identified optical counterparts are X-ray pulsars. Pulse
periods cover the range 1--10$^3$ s. On longer time scales (months to
years) the variability is also apparent in the optical and infrared bands
and it is attributed to structural changes of the circumstellar disc.  The
long-term X-ray variability of the transient Be/X-ray binaries is
characterised by two type of outbursting activity 

\begin{itemize}

\item Type I outbursts. These are regular and (quasi)periodic outbursts,
normally peaking at or close to periastron passage of the neutron star.
They are short-lived, i.e., tend to cover a relatively small fraction of
the orbital period (typically 0.2-0.3 $P_{\rm orb}$). The X-ray flux
increases by about one or two order of magnitude with respect to the pre-outburst
state, reaching peak luminosities $L_x \leq 10^{37}$ erg s$^{-1}$.

\item Type II outbursts represent major increases of the X-ray flux,
$10^{3}-10^{4}$ times that at quiescence. They reach the Eddington
luminosity for a neutron star ($\sim 10^{38}$ erg s$^{-1}$) and become the
brightest objects of the X-ray sky. They do not show any preferred orbital
phase and last for a large fraction of an orbital period or even for
several orbital periods. The formation of an accretion disc during Type II
outbursts  \citep{kris83,motc91,haya04,wils08} may occur. The discovery of
quasi-periodic oscillations in some systems \citep{ange89,take94,fing96} would
support this scenario. The presence of an accretion disk also helps explain
the large and steady spin-up rates seen during the giant outbursts, which
are difficult to account for by means of direct accretion.

\end{itemize}

In this work we present a systematic study of the aperiodic variability of
four Be/X-ray binaries during type II outbursts. First, we give a summary
of the relevant information of the sources. Sect.~\ref{obs} summarises the
observations and the characteristics of the instruments used in the
analysis. Sect.~\ref{red} refers to the reduction of the data and describes
how the CD/HID and  power spectra were obtained. A detailed description of
the results is given in Sect.~\ref{res}. This section is divided into three
subsections describing the outburst profiles, the colour analysis and the
aperiodic variability. Our interpretation of the results and the
implications of this work are discussed in Sect.~\ref{disc}.    Final
conclusions are drawn in Sect.~\ref{con}.

\begin{table*}
\caption{Observations and definition of source states.}             
\label{reg}      
\centering          
\begin{tabular}{lcccccccc}
\hline\hline
Source&Branch$^a$ &Time range	&Proposal &On-source	&$<SC>^b$ &$<HC>^b$&Flux$^{c}$	&Luminosity$^{d}$ 	 \\
state &	&MJD	     	&ID	&time (s) 	& 	&	&		&	\\
\hline \hline \noalign{\smallskip}
\multicolumn{9}{c}{{\bf \vcas}}\\
\hline \noalign{\smallskip}
Rise1	&DB	&53254.0676-53256.4660 &90089  &3872	&2.08$\pm$0.05  &0.71$\pm$0.03  &0.95	&7.5  \\
Rise2	&DB  	&53260.1495-53265.0921 &90089  &3536	&1.91$\pm$0.05	&0.81$\pm$0.04  &1.5	&11.8   \\
Peak	&DB  	&53267.1686-53272.9721 &90089  &29744   &1.87$\pm$0.01  &0.89$\pm$0.01  &1.4	&11.0   \\
Decay1	&DB	&53274.8282-53280.7687 &90089  &9712	&1.95$\pm$0.02  &0.85$\pm$0.02  &1.0	&7.9  \\
Decay2	&DB	&53278.6230-53280.7687 &90089  &6128    &2.03$\pm$0.04  &0.70$\pm$0.06  &0.8	&6.0  \\
Decay3	&DB	&53282.5715-53282.6673 &90014  &5536	&2.09$\pm$0.01	&0.62$\pm$0.01	&0.6	&5.0 \\	
Decay4	&HB	&53284.6063-53304.5730 &90014  &32500	&1.99$\pm$0.04	&0.57$\pm$0.03	&0.11	&0.9  \\
\hline \noalign{\smallskip}
\multicolumn{9}{c}{{\bf \ks}}\\
\hline \noalign{\smallskip}
Rise0	&HB	&51874.1502-51878.3595	&50425	&16720	&1.57$\pm$0.07	&1.24$\pm$0.02	&0.01	&0.12\\
Rise1	&DB	&51883.8286-51892.2963	&50425	&14060	&1.68$\pm$0.04	&1.13$\pm$0.05	&0.09	&1.0 \\
Rise2	&DB	&51896.1987-51910.9382	&50425	&17744	&1.55$\pm$0.03	&1.14$\pm$0.02	&0.25	&2.3 \\
Rise3	&DB	&51914.1750-51932.3484	&50425	&10080	&1.49$\pm$0.03	&1.16$\pm$0.03	&0.36	&4.3 \\
Peak	&DB	&51941.6795-51968.5845	&50425	&23344	&1.40$\pm$0.02	&1.20$\pm$0.02	&0.53	&6.3 \\
Decay1	&DB	&51970.2073-51990.1352	&60402	&15536	&1.51$\pm$0.03	&1.19$\pm$0.02	&0.39	&4.7 \\
Decay2	&DB	&51992.1719-52010.8263	&60402	&9888	&1.59$\pm$0.02	&1.16$\pm$0.02	&0.22	&2.6 \\
Decay3	&DB	&52014.3856-52050.1873	&60402	&9984	&1.65$\pm$0.04	&1.05$\pm$0.07	&0.11	&1.3 \\
Decay4	&HB	&52052.9510-52078.1324	&60402	&14720	&1.32$\pm$0.08	&1.11$\pm$0.04	&0.008	&0.1 \\
\hline \noalign{\smallskip}
\multicolumn{9}{c}{{\bf \exo}}\\
\hline \noalign{\smallskip}
Rise1	&DB	&53908.6156-53924.7247	&91089 \& 92067&43200	&1.39$\pm$0.07	&1.09$\pm$0.02	&1.4	&8.4	\\
Rise2	&DB	&53926.6282-53943.2576	&91089	&26512	&1.26$\pm$0.02	&1.12$\pm$0.01	&2.1	&12.7	\\
Peak	&DB	&53944.0532-53975.4312	&91089	&35536	&1.23$\pm$0.02	&1.13$\pm$0.02	&2.4	&14.5	\\
Decay1	&DB	&53976.1186-53998.6782	&91089	&25168	&1.32$\pm$0.05	&1.09$\pm$0.03	&1.6	&9.7	\\
Decay2	&DB	&53999.5902-54015.4534	&92067	&8656	&1.47$\pm$0.05	&0.99$\pm$0.02	&0.57	&3.4	\\
Decay3	&DB	&54016.5928-54030.9436	&92067	&10912	&1.43$\pm$0.03	&1.02$\pm$0.04	&0.11	&0.69	\\
Decay4	&HB	&54031.7791-54047.9337	&92067	&16144	&1.41$\pm$0.03	&0.96$\pm$0.03	&0.06	&0.37	\\
\hline \noalign{\smallskip}
\multicolumn{9}{c}{{\bf \bq}}\\
\hline \noalign{\smallskip}
Rise1	&DB  	&53340.2869-53346.7436 &90089  	&48752	&2.18$\pm$0.02  &0.89$\pm$0.03  &2.3	&13.5  \\
Rise2	&DB  	&53352.7706-53361.1469 &90089  	&32720	&2.06$\pm$0.03  &0.73$\pm$0.02  &5.0	&29.3  \\
Peak	&DB	&53364.9113-53369.5965 &90089 \& 90427&13040 &1.95$\pm$0.02  &0.66$\pm$0.01  &5.5&32.2  \\
Decay1	&DB	&53374.0267-53381.6610 &90427 \& 90014&23504 &2.04$\pm$0.02  &0.74$\pm$0.01  &4.2&24.6  \\
Decay2	&DB	&53385.8724-53395.4413 &90014	&38672	&2.11$\pm$0.03  &0.83$\pm$0.03  &2.6	&15.2 \\
Decay3	&DB	&53398.4560-53414.0726 &90014	&21296	&2.20$\pm$0.04  &0.94$\pm$0.03  &1.5	&8.8 \\
Decay4	&HB	&53416.0726-53424.4139 &90014	&20160	&2.05$\pm$0.15  &1.13$\pm$0.05  &0.29	&1.7 \\
\hline\hline  \noalign{\smallskip}
\multicolumn{9}{l}{$a$: HB: Low-intensity (Horizontal) Branch; DB: high-intensity (Diagonal) branch}\\
\multicolumn{9}{l}{$b$: Normalised to the Crab. Average Crab values are
$<SC>_{\rm Crab}=0.34$ and $<HC>_{\rm Crab}=0.47$} \\
\multicolumn{9}{l}{$c$: $\times 10^{-8}$ erg cm$^{-2}$ s$^{-1}$ in the 3--30 keV band} \\
\multicolumn{9}{l}{$d$: $\times 10^{37}$ erg s$^{-1}$ in the 3--30 keV band} \\
\end{tabular}
\end{table*}

   \begin{figure*}
   \centering
   \includegraphics[width=16cm]{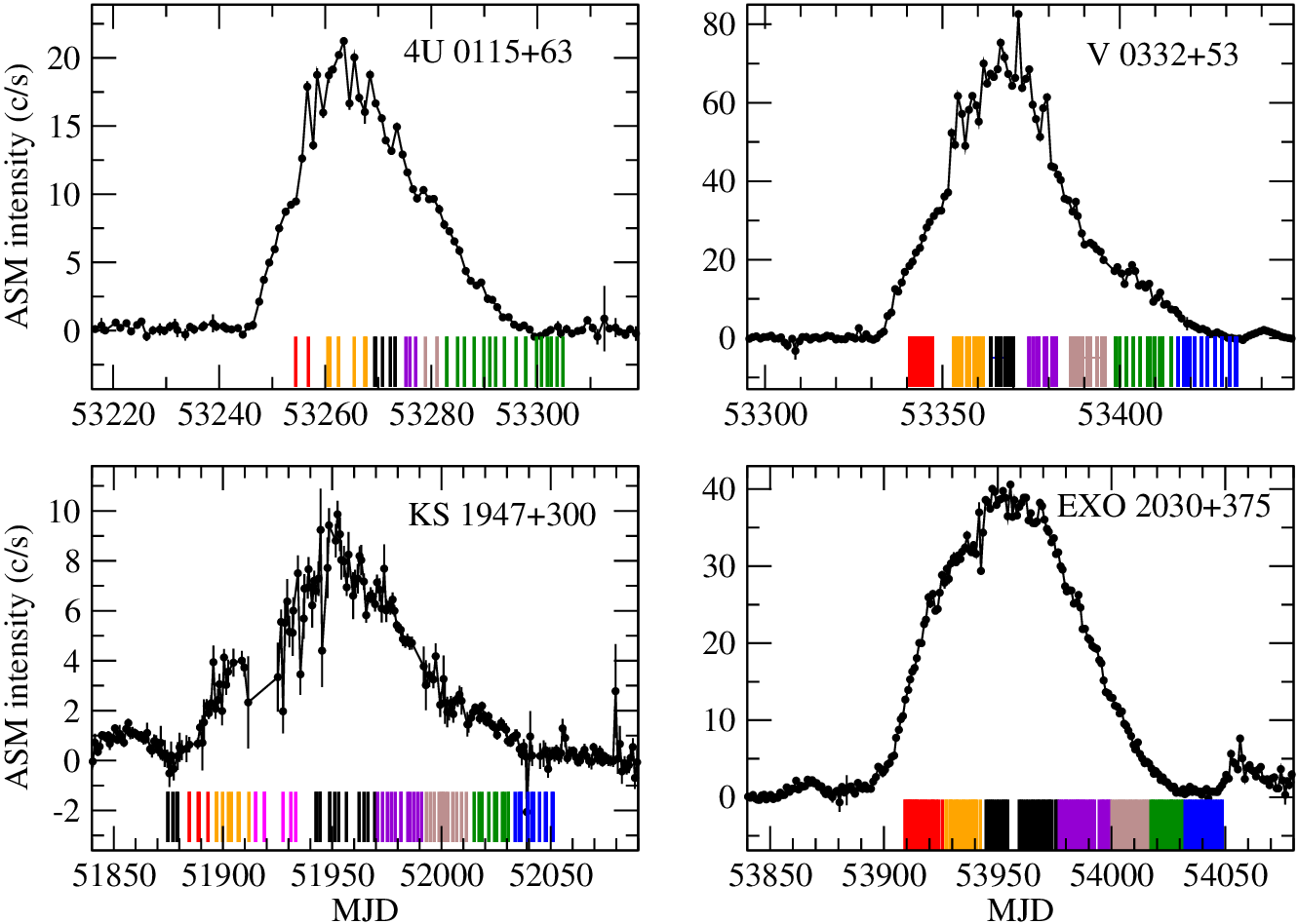} 
      \caption{The outburst profiles as seen by ASM RXTE. Vertical
      lines mark the time of the RXTE PCA observations. Different colours
      distinguish the spectral regions. Bin size is 1 day. 1 Crab corresponds 
      to $\sim$75 c/s. {\em See the electronic edition of the Journal for a 
      colour version of this figure}. 
        }
         \label{outbprof}
   \end{figure*}
\begin{figure*}
\begin{tabular}{cc}
\includegraphics[width=8cm]{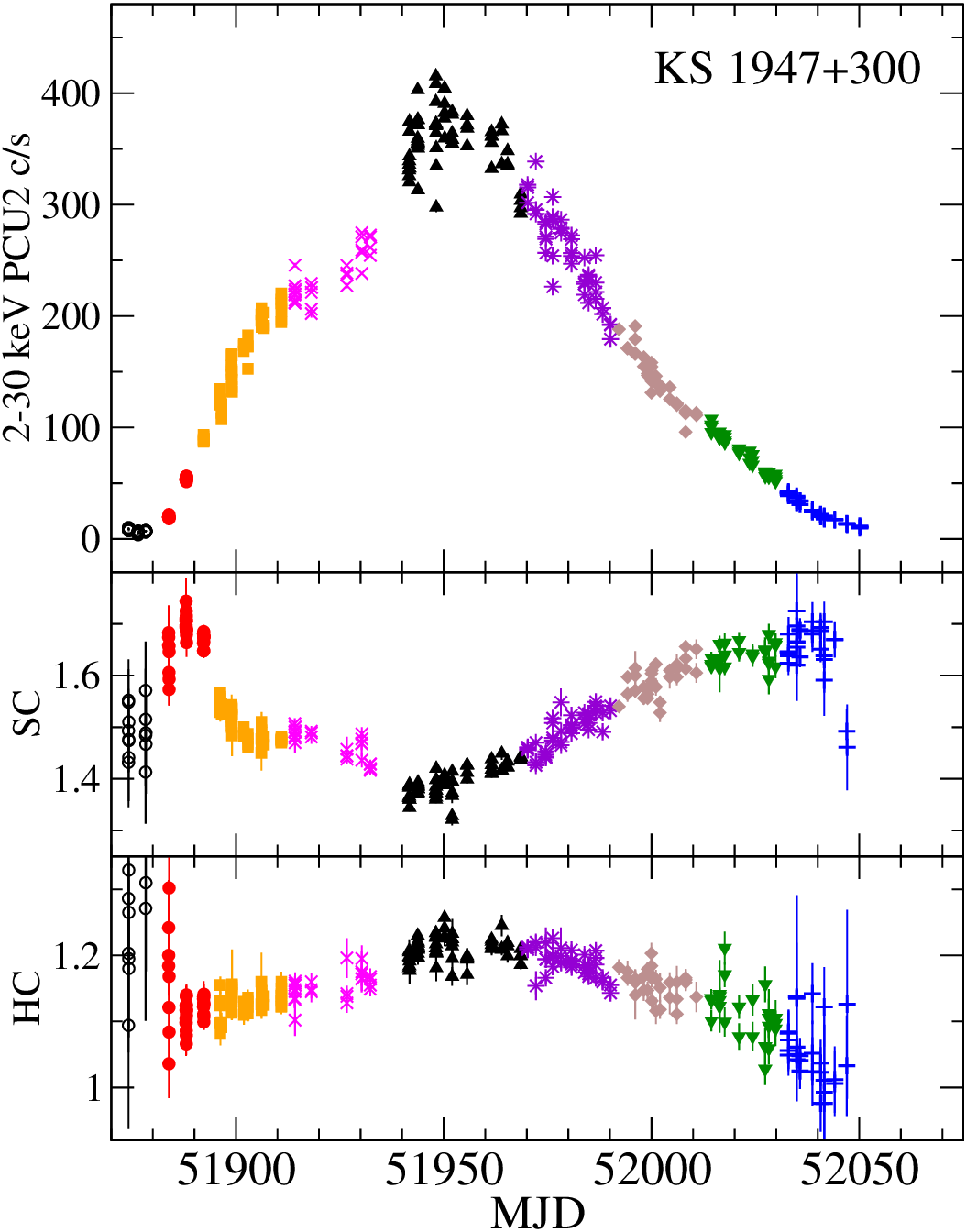} &
\includegraphics[width=8cm]{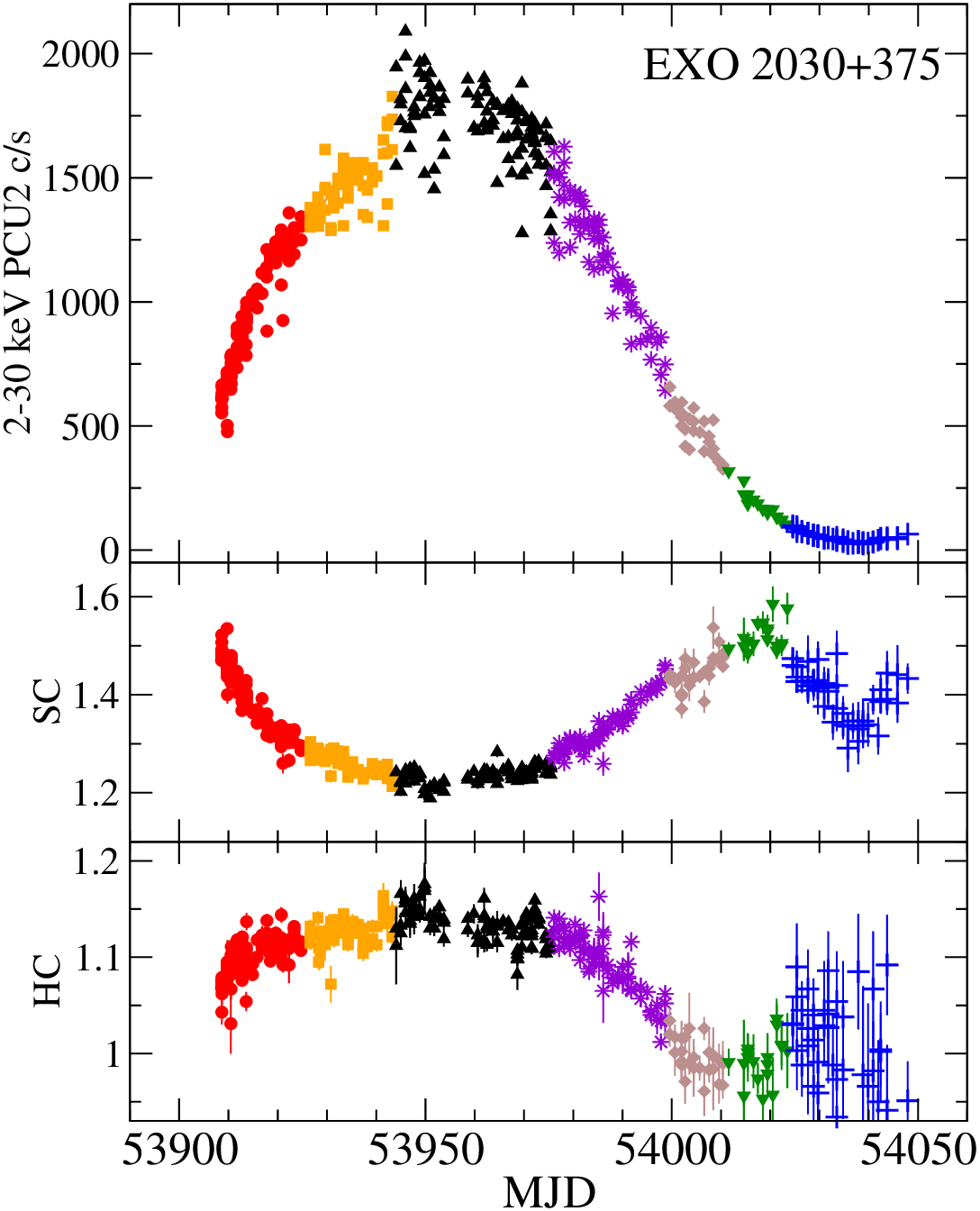} \\
\includegraphics[width=8cm]{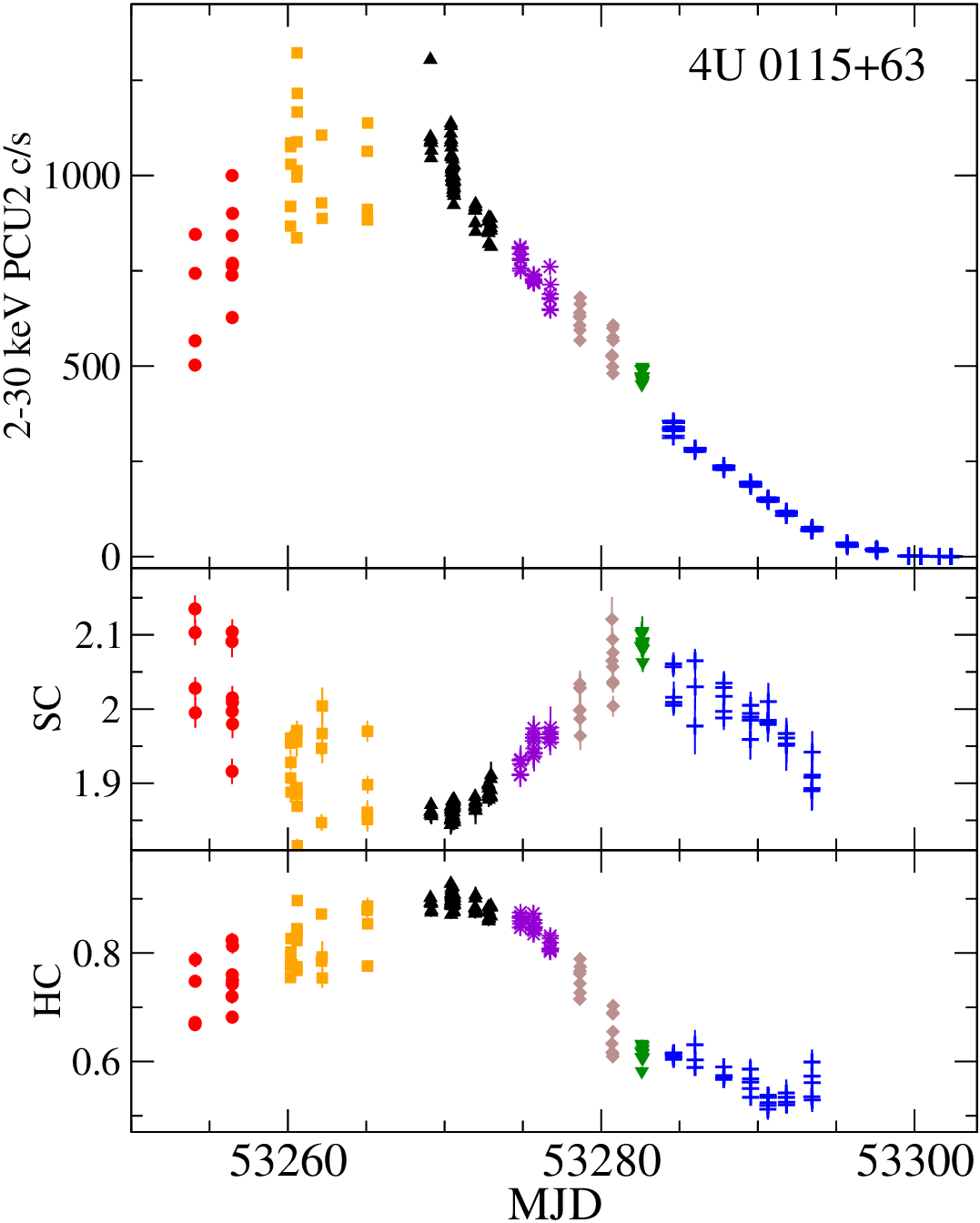} &
\includegraphics[width=8cm]{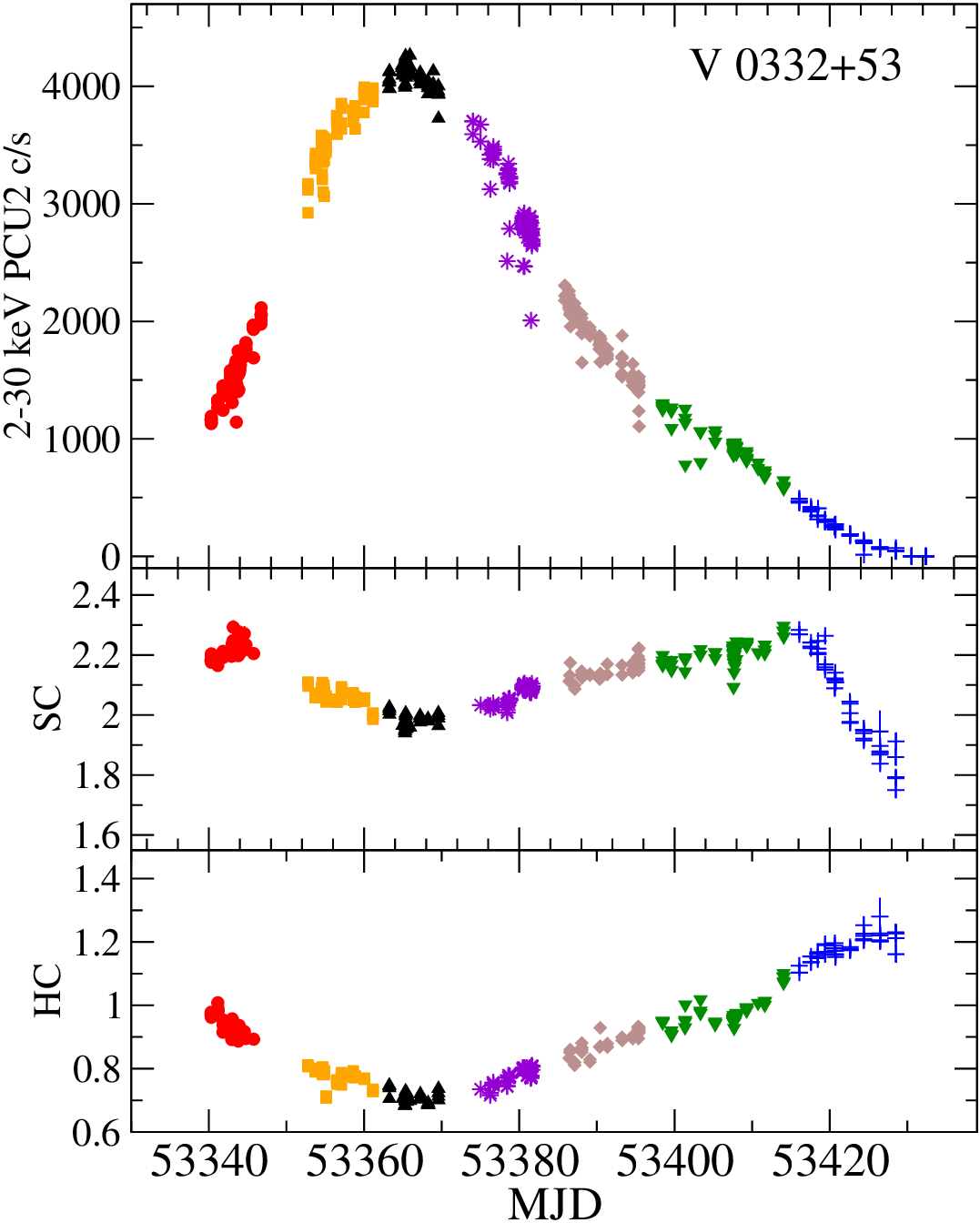} \\
\end{tabular}
\caption[]{PCA light curves and hardness ratio evolution. Each point represents 
a 512-s integration. Symbols are as follow: pre-outburst (open black
circles, only in \ks), rise1 (red circles), rise2
(orange squares), rise3 (pink crosses, only in \ks), peak (black triangle-up), 
decay1 (violet stars), decay2 (brown diamonds), decay3 (green triangle-down) 
and decay4 (blue pluses).
{\em See the electronic edition of the Journal for a colour version of 
this figure}.}
\label{rate_hr}
\end{figure*}

\section{The sources}
\subsection{4U 0115+63}

\vcas\ is one of the most active and best studied Be/X-ray transients as it
was one of the first Be/X-ray binaries to be discovered. The oldest
available X-ray observations date back to August 1969 when the {\em Vela
5B} satellite detected the source as three small outbursts separated by 180
days \citep{whit89}. Since then about 15 outbursts have been reported
\citep{reta07}. During the RXTE's life time 3 major outbursts have been
detected: in 1999, 2000 and 2004. Here we analyse observations from the
2004 event. The neutron star orbits a B0.2Ve in a moderately eccentric orbit 
with $P_{\rm orb}=24.3$ d, $e=0.34$ and $a_x \sin i=140.1$ lt-s
\citep{rapp78} and rotates with $P_{\rm spin}=3.6$ s, as inferred from the
pulsed X-ray emission. The distance to the source is estimated to be
$\sim$8 kpc \citep{reta07}.

The X-ray spectrum shows up to five cyclotron resonance scattering features
(fundamental and four harmonics) and constitutes the accreting neutron star
with the highest number of cyclotron lines \citep{hein04}. The optical and
infrared emission is characterised by cyclic changes with  a period of
$\sim$ 5 years. Another peculiarity is that  X-ray outbursts in \vcas\
appear to come in pairs, i.e.,  two in every cycle. However, sometimes the
second outburst is missing \citep{reta07}. mHZ QPOs were detected during
the 1999 and 2004 outbursts \citep{hein99,cobu04}.

\subsection{\ks}

The Be/X-ray binary \ks\ was first detected in the X-ray band on June 8,
1989 by the TTM coded mask X-ray spectrograph on the Kvant module of the
Mir space station \citep{boro90} with a peak flux of 70 mCrab. About 35
days later the flux was at the limit of sensitivity of the instrument with
a 3$\sigma$ upper limit of 10 mCrab.  X-ray pulsations with a pulse period
of 18.7 s were found in the X-ray flux of GRO J1948+32 in April 1994 by the
all-sky monitor BATSE aboard CGRO \citep{chak95}. The source reached a peak
pulse flux of 50 mCrab in the energy range 20-75 keV. About 25 days later
the flux decayed bellow the detection threshold of BATSE. Given the very
large error box of GRO J1948+32 no connection with this source and \ks\ was
realised. The source remained in quiescence until the beginning of 2001 
\citep{galo04}. In this occasion the source reached a maximum flux of
120mCrab, the largest in its history. This outburst is the object of the
present work. Prior to this major outburst, weak  emission ($\sim$ 20
mCrab) and X-ray pulsations (18.7579$\pm$0.0005 s) from \ks\ had been
detected in 2000 November \citep{levi00,swan00}. The coincidence in the
value of the pulse period and the location of \ks\ inside the GRO J1948+32
error circle suggest that the two sources are in fact the same object. 
Pulse timing analysis allowed to solve for the orbital parameters resulting
in orbital period $P_{\rm orb}=40.415\pm0.010$ d, eccentricity
$e=0.033\pm0.013$ and projected radius $a_x \sin i=137\pm3$ lt-s
\citep{galo04}.  A broad-band spectral study of \ks\ using {\it BeppoSAX}
observations \citep{naik06} showed that the energy spectrum in the 0.1--100 keV energy
band has three components: a Comptonized component, a $\sim$0.6 keV
blackbody component, and a narrow, weak iron emission line at 6.7 keV with
a low column density of material in the line of sight.  Although the
optical counterpart to \ks\ was correctly identified soon after the Kvant
detection \citep{gora91,gran91} its identification as a Be/X-ray binary was
suggested by \citet{negu03}. \ks\ is associated with a moderately reddened
V=14.2 B0Ve star located at $\sim$ 10 kpc.

\begin{figure*}
\begin{tabular}{cc}
\includegraphics[width=8cm]{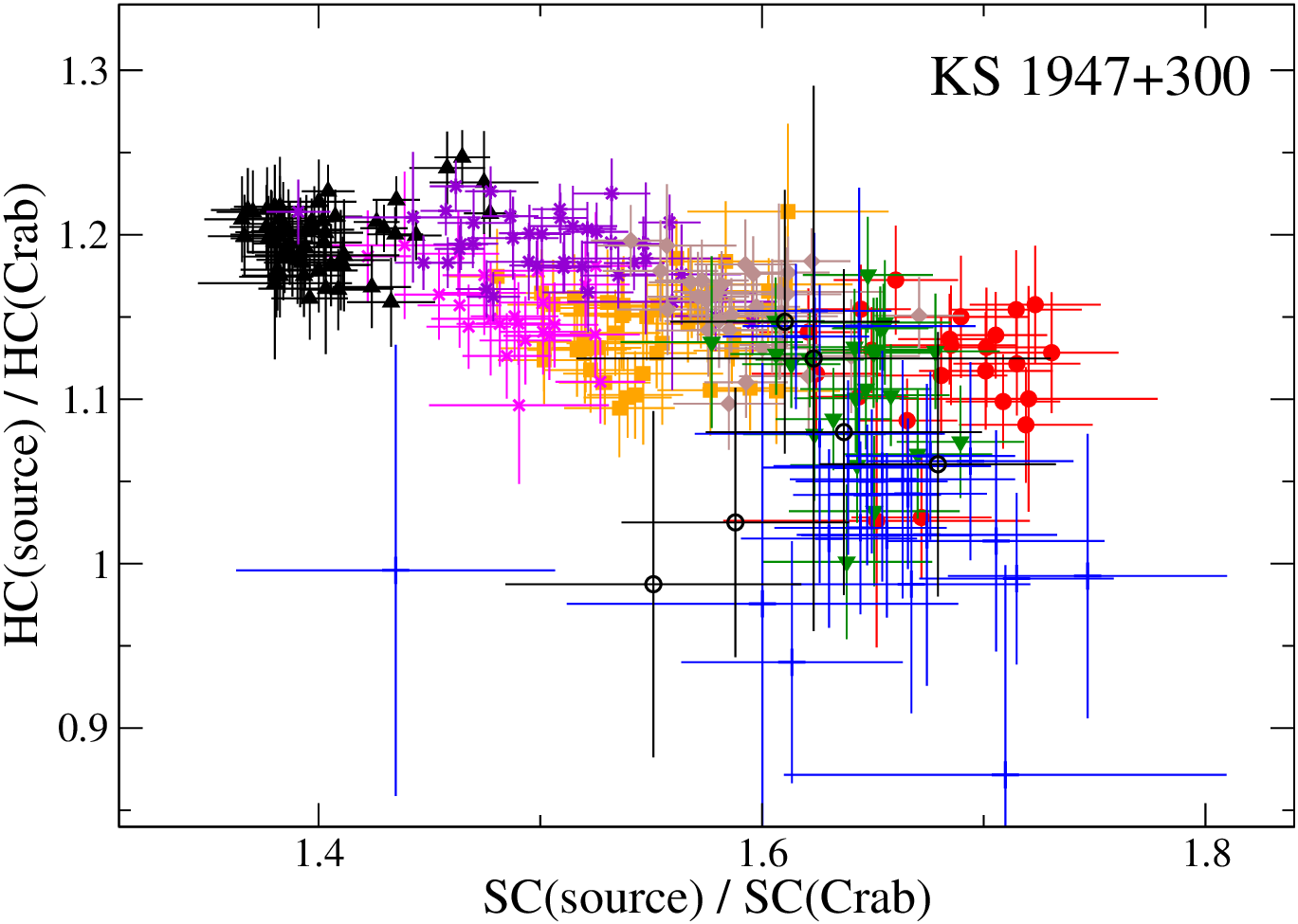} &
\includegraphics[width=8cm]{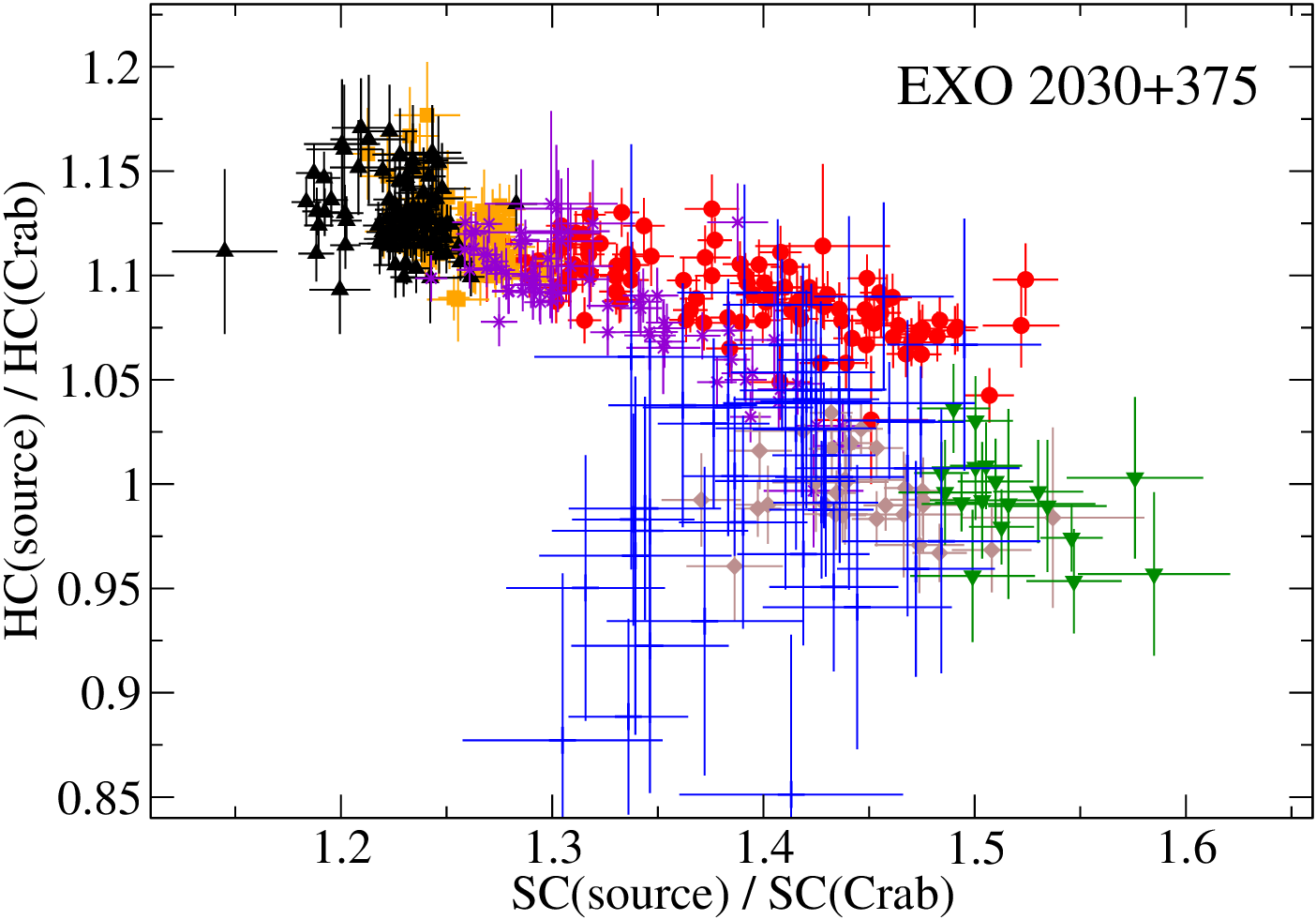} \\ 
\includegraphics[width=8cm]{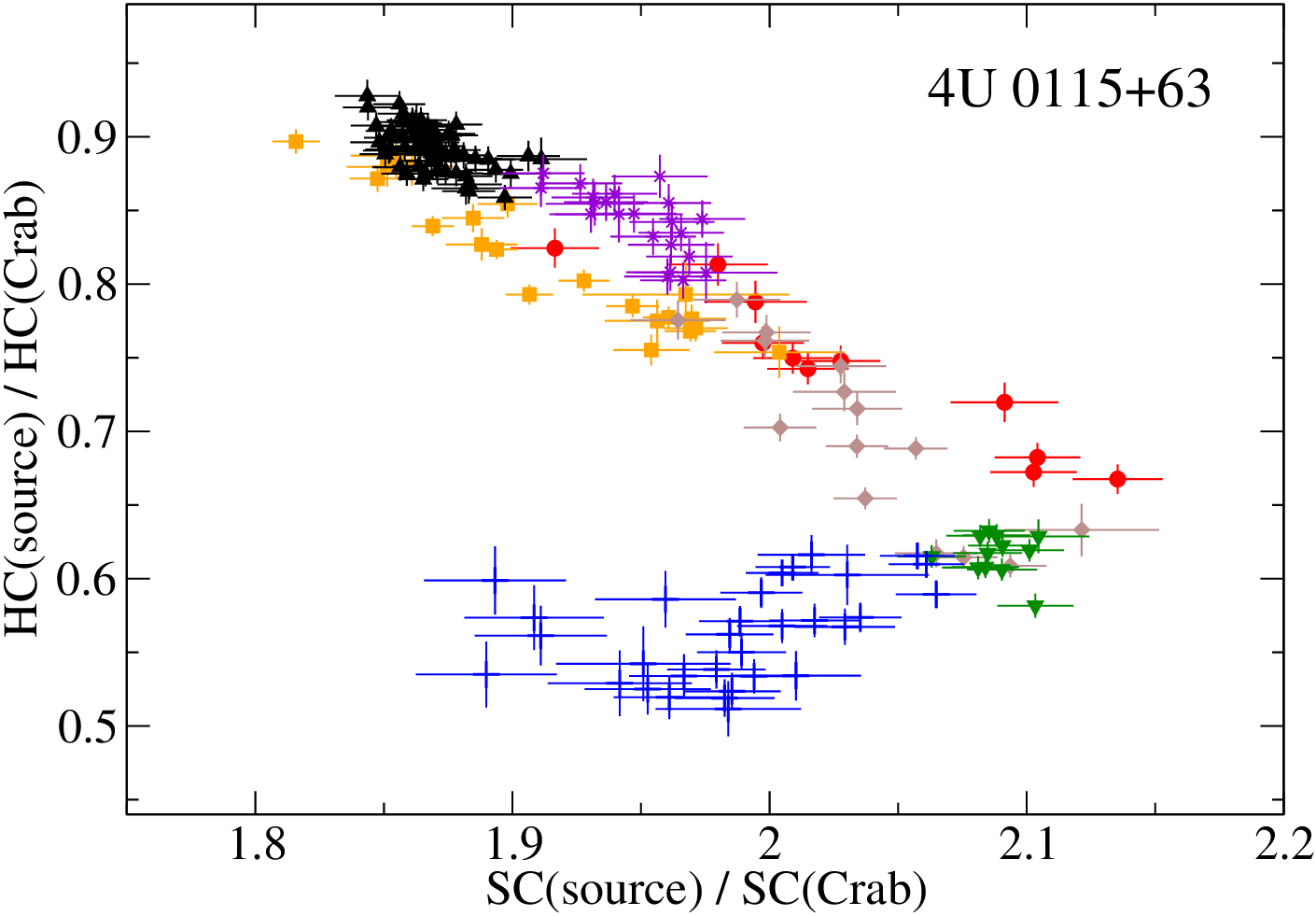} &
\includegraphics[width=8cm]{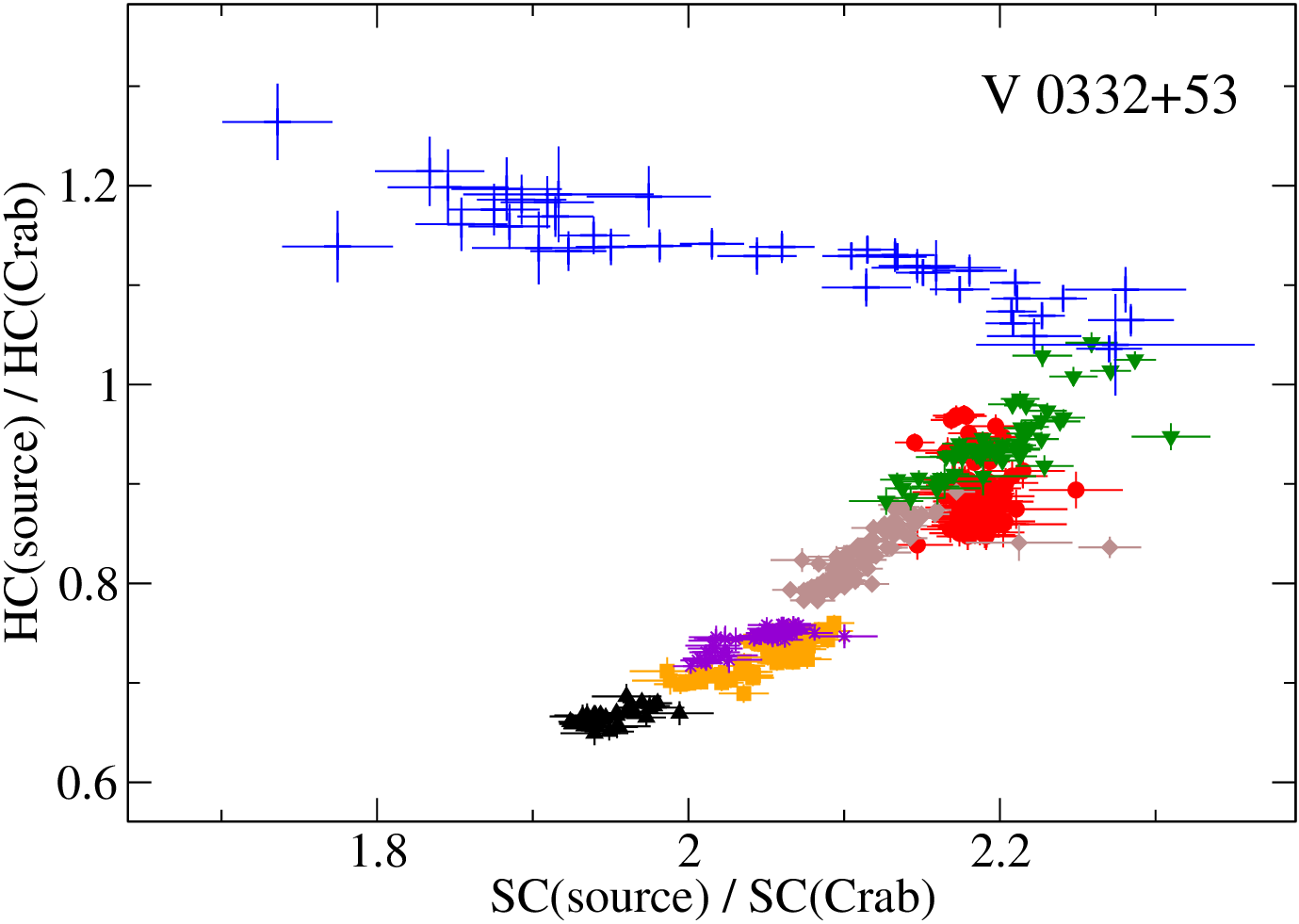} \\         
\end{tabular}
\caption[]{Colour-colour diagrams of the four sources
investigated in this work. Each point represents a 512-s integration. 
SC=7-10 keV/4-7 keV, HC=15-30 kev/10-15 keV. Symbols are as follows: 
rise0 (open black circles, only in \ks), rise1 (red circles), rise2
(orange squares), rise3 (pink crosses, only in \ks), peak (black triangle-up), 
decay1 (violet stars), decay2 (brown diamonds), decay3 (green triangle-down) 
and decay4 (blue pluses). {\em See the electronic edition 
of the Journal for a colour version of this figure}.}
\label{cd}
\end{figure*}
\begin{figure*}
\begin{tabular}{cc}
\includegraphics[width=8cm]{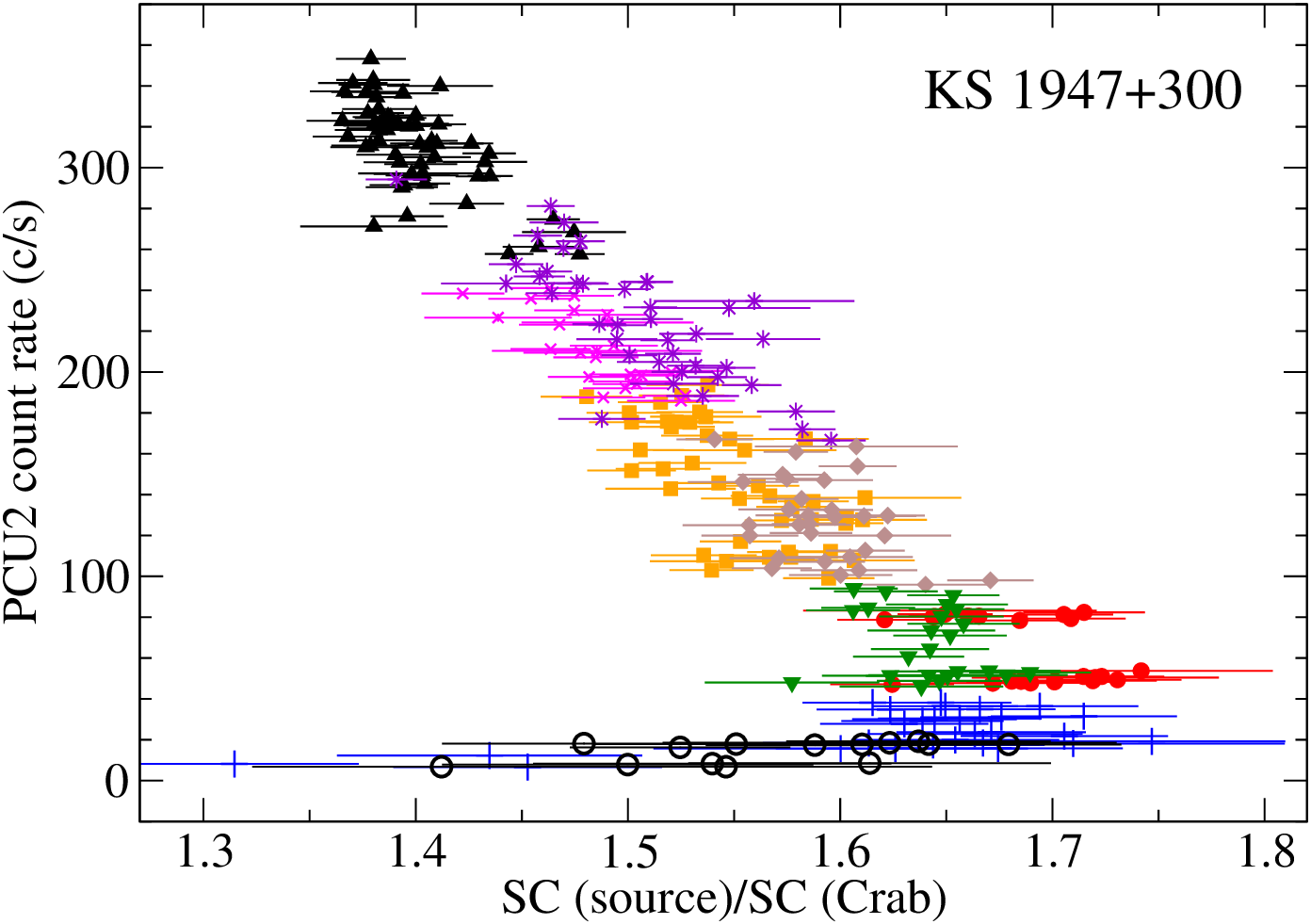} &
\includegraphics[width=8cm]{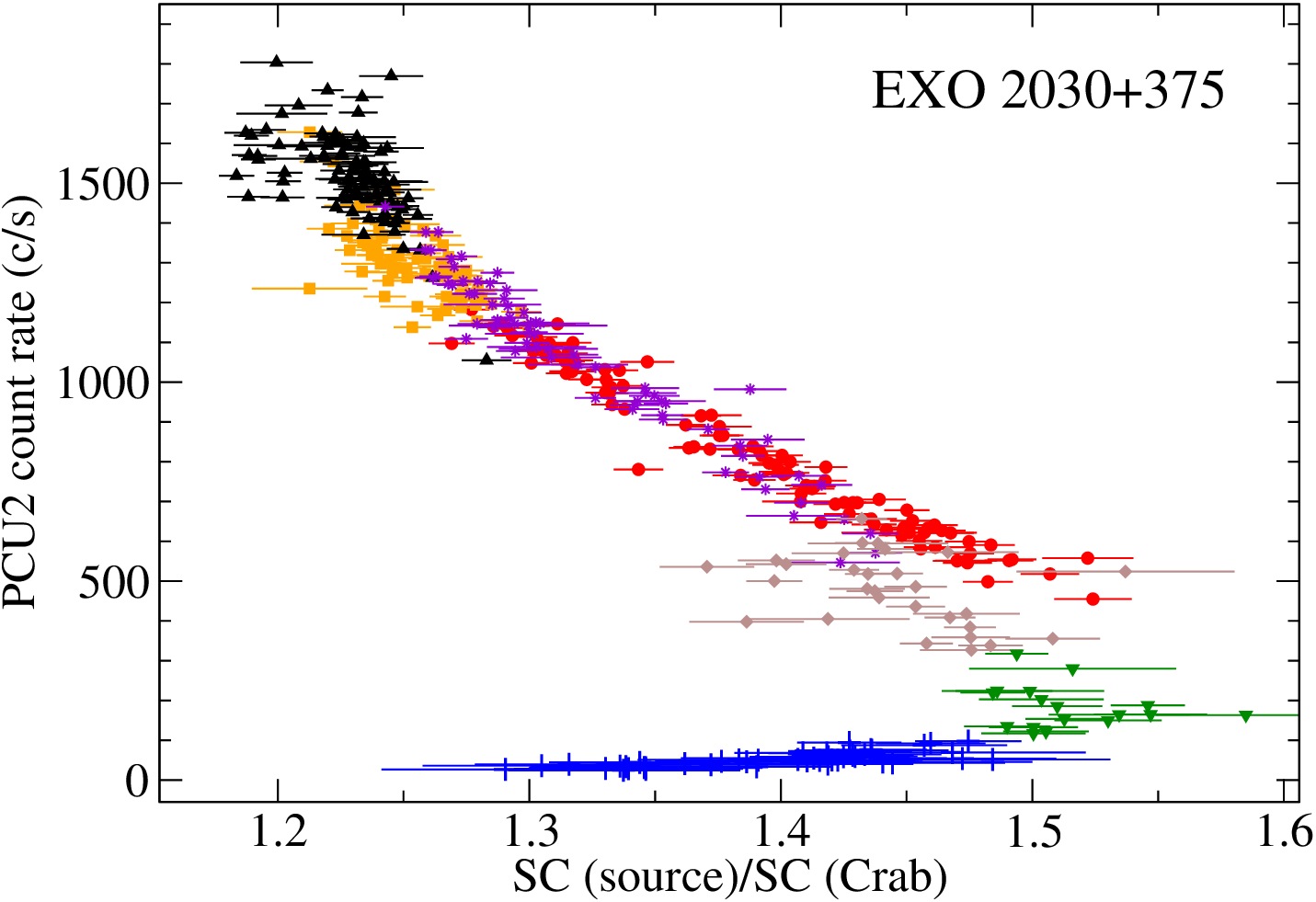}\\
\includegraphics[width=8cm]{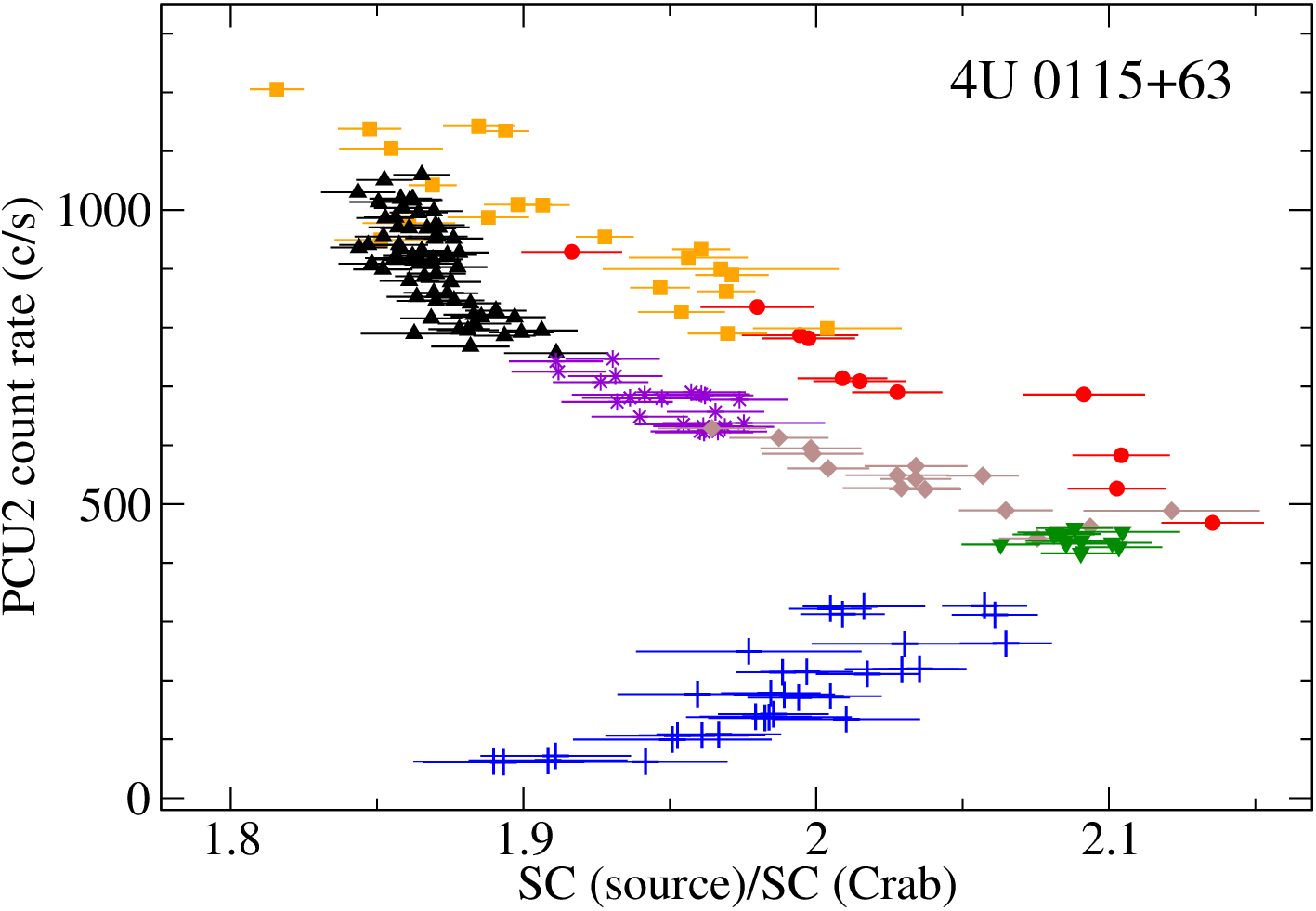} &
\includegraphics[width=8cm]{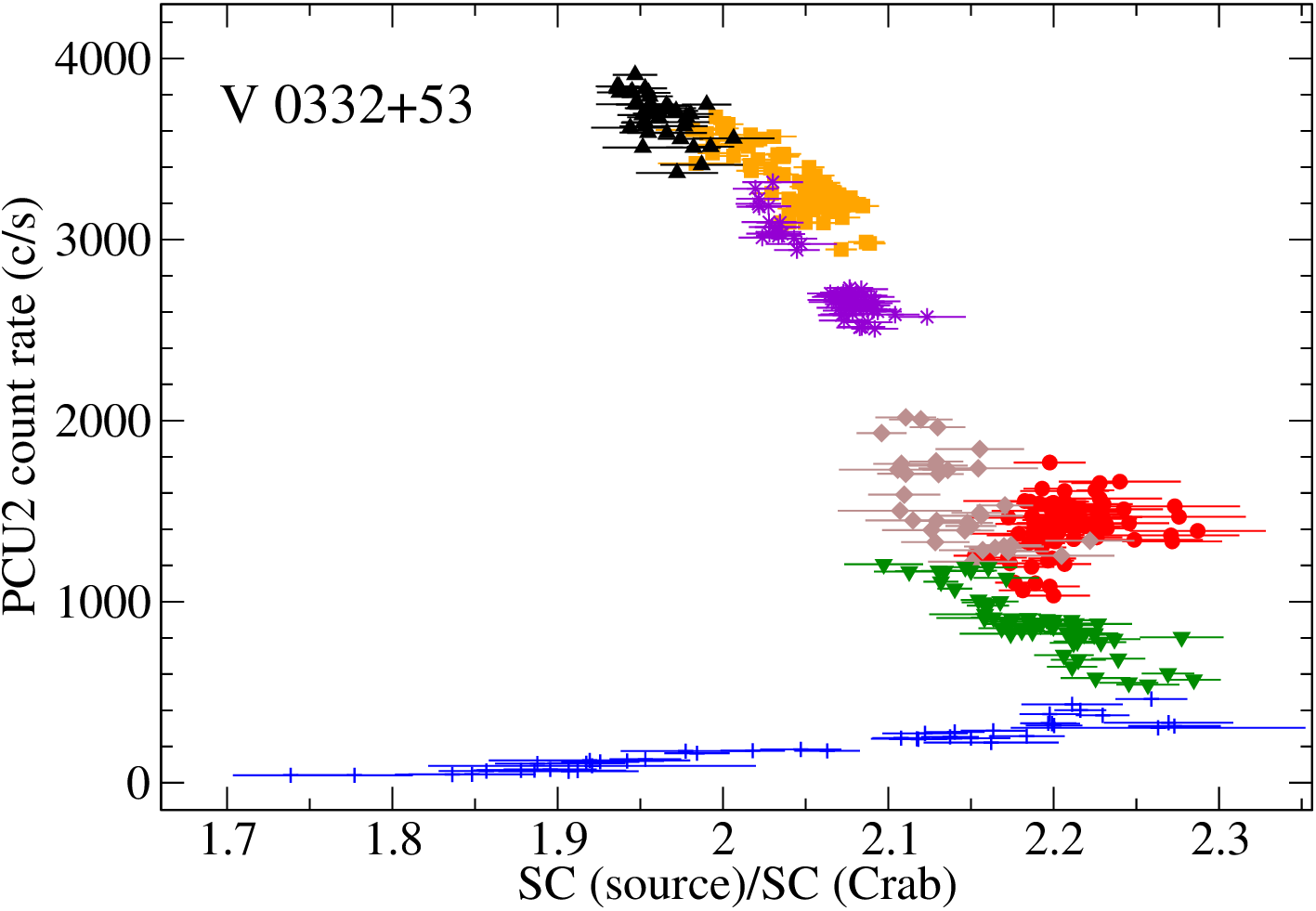}\\
\end{tabular}
\caption[]{Hardness-intensity diagrams. Each point represents a 512-s integration.
Intensity corresponds to the energy range 4-30 keV. Symbols are as follows:
rise0 (open black circles, only in \ks), rise1 (red circles), rise2
(orange squares), rise3 (pink crosses, only in \ks), peak (black triangle-up), 
decay1 (violet stars), decay2 (brown diamonds), decay3 (green triangle-down) 
and decay4 (blue pluses).  
{\em See the electronic edition of the Journal for a colour version of this 
figure}.}
\label{hid}
\end{figure*}

\subsection{\exo}

\exo\ was discovered  by EXOSAT in 1985 \citep{parm89}.  Since then it has
been extensively observed in the X/$\gamma$-ray band by various missions:
EXOSAT \citep{reyn93},  ROSAT \citep{mavr94},  RXTE \citep{reig98,reig99},
CGRO \citep{stol99}, INTEGRAL \citep{came05} and SWIFT \citep{kloc07}.
\exo\ is the prototype of X-ray variability in Be/X-ray binaries. It shows
a regular pattern of type I outbursts ($L_{\rm x}\approx 10^{36}$ erg
s$^{-1}$) during each periastron passage \citep{wils02}. In addition, \exo\
has shown two major outbursts  ($L_{\rm x}\simmore 10^{38}$ erg s$^{-1}$):
the first one in 1985 that led to its discovery as an X-ray transient and
the second one in 2006. In this work we analyse RXTE observations of the
latest event.

\exo\ contains a rotating neutron star ($P_{\rm spin}=41.7$ s) orbiting a
Be primary \citep{motc87,coe88,reigal98} in a moderately eccentric
($e=0.4$) wide ($P_{\rm orb}=46$ d) orbit and is located at about 7.1 kpc
\citep{wils02}. Quasi-periodic oscillations with a frequency of 0.2 Hz were
detected from this source when the luminosity was close to its maximum
during the 1985 outburst \citep{ange89}. In common with other high-mass
X-ray binaries, the continuum spectral shape in the range 1-30 keV  can be
represented by a power law with an exponential cutoff \citep{reyn93}. A
blackbody component has been reported to give good fits at very high
($L_{\rm x}\sim 10^{38}$ erg s$^{-1}$) luminosity \citep{sun94} and
relatively low ($L_{\rm x}\approx 10^{36}$ erg s$^{-1}$) luminosity
\citep{reig99}.

\subsection{\bq}

A similar analysis to the one presented here has already been published in
\citet{reig06}. For the sake of clarity and in order to apply the same
reduction procedure to all sources, we have re-analysed the observations in
\citet{reig06}. Note also that the present work includes data during the
rise of the outburst that were not previously analysed and that the
frequency range of the power spectra is two orders of magnitude larger than
in \citet{reig06}.

\bq\ is a hard X-ray transient that spends most of its life in a quiescent
X-ray state. After more than 15 years of quiescence the Be/X-ray binary
\bq\ underwent a giant outburst in December 2004. The outburst was
predicted about 9 months before its occurrence from the brightening of the
optical companion \citep{gora04}. Prior to the event reported here, \bq\ had
went into outburst is three other occasions. The 1973 outburst led to its
discovery as a bright X-ray source by the {\em Vela 5B} satellite
\citep{terr84}. In 1983 it reappeared in the form of three small
outbursts \citep{tana83}. EXOSAT observations of this activity period
resulted in the discovery of X-ray pulsations with $P_{\rm spin}=4.4$ s and
the determination of the orbital parameters, $P_{\rm orb}=34.25$ d and
$e=0.31$ \citep{stel85}. In 1989 {\em Ginga} detected \bq\ again
and allowed the discovery of a cyclotron resonant scattering feature at
28.5 keV \citep{maki90} and QPOs at 0.051 Hz \citep{take94}. The optical
counterpart to \bq\ is an O8-9Ve star at a distance of $\sim 7$ kpc,
showing H$\alpha$ in emission and strong and variable infrared emission
\citep{bern84,corb86,coe87,negu99}.

\section{Observations}
\label{obs}

We have analysed data obtained by the Proportional Counter Array (PCA) and
the All Sky Monitor (ASM) onboard the Rossi X-ray Timing Explorer (RXTE).
The ASM data consist of daily flux averages in the energy range 1.3-12.1
keV. The PCA covers the lower part of the energy range 2--60 keV, and 
consists of five identical coaligned gas-filled proportional units.
Due to RXTE's low-Earth orbit, the data consist of a number of contiguous
data intervals (typically 1 hr long) interspersed with observational gaps
produced by Earth occultations of the source and passages of the satellite
through the South Atlantic Anomaly. We shall refer to each of these data
intervals as  "pointings". Data taken during satellite slews, passage
through the South Atlantic Anomaly and Earth occultation were removed.

During the RXTE life time the response of the detectors varied due to
ageing. Also, gain changes are applied occasionally, making the channel
boundaries for a given energy range change with time, and also slightly
affecting the effective areas of the detectors. Each gain change is the
start of a new "gain epoch". To minimize this effect, data from the
same "epoch" was used. All the data analysed in this work correspond to
"epoch" 5. In addition, the source X-ray colours were normalised to those
of the Crab that are closest in time.  Table~\ref{reg} gives relevant
information about the observations.

\section{Data reduction and analysis}
\label{red}

In this section we describe the different tools (CD/HID and power
spectra) used in the data analysis. ASM light curves were obtained to study
the outburst profiles. PCA data can be collected and telemetered to the
ground in many different ways depending on the intensity of the source and
the spectral and timing resolution desired. In this work we used {\em
Standard2} to obtain colour-colour and hardness-intensity diagrams and
time-average energy spectra. In addition, various high-time resolution
modes ({\em good\_xenon, Event\_125$\mu$s\_64M\_0\_1s and
SB\_62$\mu$s\_0\_49}) were used to extract the light curves with which the
power spectra were produced.

\subsection{Outburst profiles}

The outburst profiles were obtained from the 1-day binned ASM light curves
and are shown in Figure~\ref{outbprof}. Table~\ref{info} gives the duration
and peak intensity of each individual outburst together with some relevant
information about the systems. The time of the PCA observations are marked
in Fig.~\ref{outbprof} with different colours indicating different spectral
regions (see the electronic edition of the Journal for a colour version of
this figure). Each vertical mark corresponds to a pointing. In the energy
range of the ASM, 1 Crab corresponds to $\sim$75 c s$^{-1}$. In
general, the PCA sampling of the outbursts is excellent. The only part of
the outburst not covered by the PCA data is the 7-10 first days, which can
be attributed to the transient nature of the sources (it takes some time to
issue an alert warning and reschedule the observations). The only source
with available data covering the entire outburst is \ks. 

\subsection{Colour-colour and hardness-intensity diagrams}
\label{col}

Colour-colour (CD) and hardness-intensity (HID) diagrams are powerful
tools that allow the investigation of spectral variability without the
assumption of any spectral model.
In order to study the evolution of the spectral shape as the
outburst evolves, the RXTE light curve was divided into different parts
depending on count rate and whether the outburst was on the rise, peak or
decay. Given the good sampling of the PCA observations, these
three phases were further subdivided into shorter intervals. Thus each
interval defines a spectral region characterised by count rate, soft and
hard colours. Table~\ref{reg} gives relevant information about each region
and source.

The CD and HID were obtained from the background-subtracted {\em Standard
2} PCA light curves in the energy ranges $c_1=4-7$ keV, $c_2=7-10$ keV, 
$c_3=10-15$ keV and $c_4=15-30$ keV. The soft colour (SC) was defined as
the ratio $c_2/c_1$ and the hard colour (HC) as the ratio  $c_4/c_3$. These
ratios are expected to be insensitive to interstellar absorption effects.
The hydrogen column densities to the systems, obtained from model fits to
the X-ray spectra, are in the range $0.6-3\times 10^{22}$ cm$^{-2}$. The
evolution with time of the PCU2 count rate and the soft and hard colours is
shown in Fig.~\ref{rate_hr}.

The CD was constructed by plotting the hard colour as function of the soft
colour (Fig.~\ref{cd}) and the HID by plotting the count rate in the 4-30
keV band as a function of one of the colours (Fig.~\ref{hid}).  To recover
the true values of the colours of the sources, the soft colour should be
multiplied by 0.34 and the hard color by 0.47 (quoted values are averages
of the Crab colours during the observations). The variation of the Crab
colours computed as the root-mean-square, i.e. the standard deviation over
the mean colour, was, on average, 2.1\% and 1.6\% for the soft and hard
colours, respectively.

To reduce the uncertainty in the individual 16-s points (default
resolution of the  {\em Standard 2} mode) the light curves were rebinned to
512-s bins. Also, data for which the resulting relative errors were larger
than 5\% were excluded. When the count rate per PCU was lower than 10 c
s$^{-1}$, relative errors up to 10\% were allowed. 

   \begin{figure}
   \centering
   \includegraphics[width=8cm]{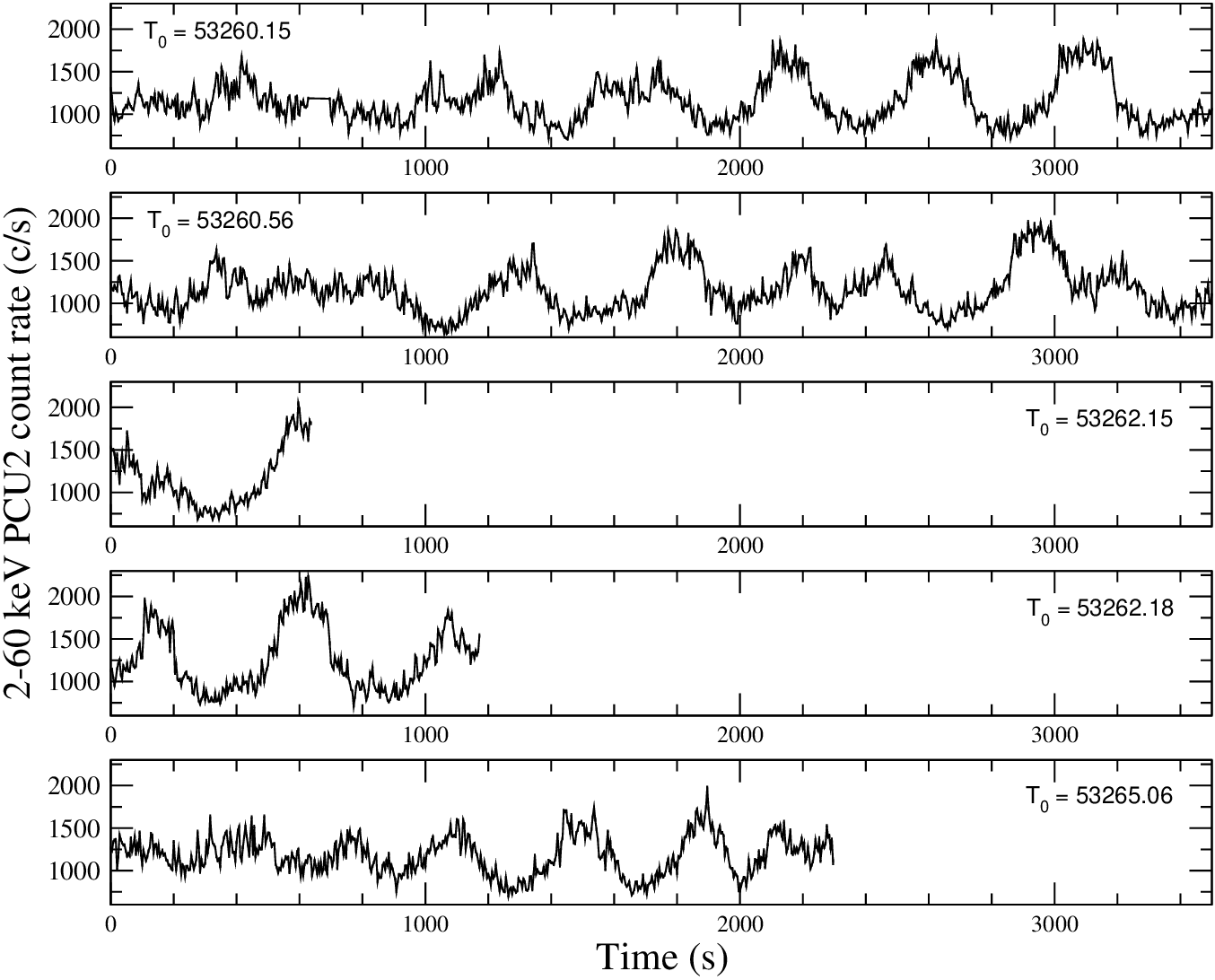}
      \caption{QPOs near the peak of the outburst in \vcas. Time zero is
      indicated in MJD in each panel. The time resolution is 4 s.}     
      \label{qpo_0115}
   \end{figure}

\begin{table*}
\caption{Power spectral parameters (centroid frequency, FWHM and $rms$) for the 
$L_b$, $L_l$ and $L_u$ components. Errors are 90\% confidence level.  The
centroid frequency of the zero-centred Lorentzian was fixed during the fit.
HB stands for horizontal branch.}             
\label{specres1}      
\centering          
\begin{tabular}{lc|ccc|ccc|cccc}
\hline\hline       
Spectral&$L_x$	&\multicolumn{3}{c}{$L_b$}					&\multicolumn{3}{c}{$L_l$}			&\multicolumn{3}{c}{$L_u$}	&	\\
region	&$\times 10^{37}$&$\nu_b$ 	&$FWHM_b$ 	 &$rms_b$	&$\nu_l$ 	&$FWHM_l$	&$rms_l$	&$\nu_u$ &$FWHM_u$	&$rms_u$&$\chi^2$/		\\
	&erg s$^{-1}$ 	&		&$10^{-3}$ (Hz)	 &(\%)		&(Hz)		&(Hz)		&(\%)		&(Hz)	&(Hz)			&(\%)	&dof$^*$		\\
\hline \hline \noalign{\smallskip}																																
\multicolumn{12}{c}{{\bf \vcas}} \\																																
\hline	\noalign{\smallskip}																															
rise1	&7.5   &0			&2.0$^{fixed}$ 	&19$\pm$3	&0		&1.6$\pm$0.3 	 &8.7$\pm$0.6	 	&--   		&-- 		&--   		&1.4/78 \\
rise2	&11.8  &0			&2.0$^{fixed}$ 	&12$\pm$2	&0		&0.87$\pm$0.06 	 &10.2$\pm$0.2  	&0.1$\pm$0.1	&4.1$\pm$0.3 	&8.2$\pm$0.3   &1.3/76 \\
peak	&11.0  &**	&**  	&**    &0		&0.78$\pm$0.08	 &9.5$\pm$0.3      	&0.4$\pm$0.3	&4.9$\pm$0.7	&4.0$\pm$0.7   &2.0/76 \\
decay1	&7.9   &**	&**  	&**	&0		&0.72$\pm$0.07	 &10.0$\pm$0.3		&1.5$\pm$0.7	&4.3$\pm$0.9	 &3.9$\pm$0.4   &1.3/76 \\
decay2	&6.0   &0			&3.0$\pm$2.0 	&13$\pm$3	&0		&0.76$\pm$0.10	&9.7$\pm$0.3   		&0   		&5.3$\pm$1.0 	&3.7$\pm$0.7   &1.9/74 \\
decay3 &5.0    &**	 &**	&**	&0		&0.73$\pm$0.04	 &11.0$\pm$0.2		&--  		&-- 		&--  		 &1.3/75 \\
decay4(HB)&0.9	&--			&--		&--		&0.10$\pm$0.01	&0.34$\pm$0.01	&20.0$\pm$0.3		&--  		&--		&-- 		  &1.1/78 \\
\hline	\noalign{\smallskip}																															
\multicolumn{12}{c}{{\bf \ks}} \\																																
\hline	\noalign{\smallskip}
rise0(HB)&0.1   &--		&--		 &--		 &0	&0.087$\pm$0.005  &35.4$\pm$0.7   &--	 &--	 &--    &2.1/83 \\																															
rise1	&0.8   &--		&--		 &--		 &0	&0.14$\pm$0.01	  &17.0$\pm$0.4   &--	 &--	 &--    &1.6/86 \\
rise2	&2.3   &0		&9$\pm$3	 &9.7$\pm$0.5     &0	 &0.40$\pm$0.04	   &14.0$\pm$0.5   &0    &2.3$\pm$0.8	  &6.3$\pm$1.2    &0.9/82 \\
rise3	&4.3   &0		&15$\pm$10	 &8.7$\pm$0.9     &0	 &0.57$\pm$0.04	   &13.2$\pm$0.2   &0    &4.0$\pm$0.8	  &6.2$\pm$0.4    &1.7/82 \\
peak	&6.3   &0		&12$\pm$2	 &9.5$\pm$0.5     &0	 &0.65$\pm$0.04	   &14.3$\pm$0.2   &1.1$\pm$0.3  &3.6$\pm$0.2  &6.5$\pm$0.5  &1.3/81 \\
decay1	&4.7   &0		&9$\pm$2	 &10.2$\pm$0.7    &0	 &0.48$\pm$0.05	   &13.8$\pm$0.5   &0   &2.7$\pm$0.4  &7.7$\pm$0.6    &1.0/82 \\
decay2	&2.6   &0		&14$\pm$3   	 &9.2$\pm$0.8     &0	 &0.61$\pm$0.04	   &14.0$\pm$0.4   &--    &--	  &--    &1.5/84 \\
decay3	&1.3   &0		&11$\pm$9	 &8.4$\pm$0.9     &0	 &0.52$\pm$0.06	   &12.0$\pm$0.6   &--    &--	  &--    &1.0/84 \\
decay4(HB)&0.01  &--		&--		&--		 &0	&0.10$\pm$0.02	  &15.8$\pm$0.6	  &--	  &--	  &--	&1.4/87 \\
\hline	\noalign{\smallskip}																															
\multicolumn{12}{c}{{\bf \exo}} \\																																
\hline	\noalign{\smallskip}																															
rise1	&8.4   &0	&44$\pm$6	&15.2$\pm$0.5   &0		&0.50$\pm$0.02   &21.7$\pm$0.3	&0		&2.9$\pm$0.2   &10.5$\pm$0.2  &1.7/81 \\
rise2	&12.7  &0	&15$\pm$6	&6.9$\pm$0.4    &0.07$\pm$0.01	&0.55$\pm$0.02   &24.2$\pm$0.2 	&0.79$\pm$0.15	&1.8$\pm$0.1   &9.8$\pm$0.7	&1.8/79 \\
peak	&14.5  &0	&3.3$\pm$2.0	&9.7$\pm$2.1    &0.20$\pm$0.01	&0.50$\pm$0.02   &20.9$\pm$0.7 	&0.79$\pm$0.15	&2.4$\pm$0.2   &12.3$\pm$1.4   &2.5/75 \\
decay1	&9.7   &0	&3.4$\pm$1.0	&10.2$\pm$1.0   &0.17$\pm$0.02	&0.64$\pm$0.09	 &22.0$\pm$1.5	&0.74$\pm$0.60  &2.3$\pm$0.8   &10.4$\pm$4.9	&1.4/76 \\
decay2	&3.4   &0	&12$\pm$5      &8.1$\pm$0.9	&0.11$\pm$0.04  &0.83$\pm$0.07   &21.7$\pm$1.3 	&0.54$\pm$0.60	&5.3$\pm$1.2  &9.2$\pm$0.9	&1.3/80 \\
decay3	&0.7	&0	&42$\pm$20	&9.2$\pm$0.8  	&0.20$\pm$0.09	&0.64$\pm$0.07	&15.6$\pm$0.7  	&0.95$\pm$0.80	&4.3$\pm$0.8	&7.8$\pm$0.4	&1.1/78 \\
decay4(HB)&0.4 &--	&--      	&--  		&0		&0.046$\pm$0.005&22.1$\pm$0.9	&--		&--		&--		&2.0/85 \\
\hline	\noalign{\smallskip}																															
\multicolumn{12}{c}{{\bf \bq}} \\																																
\hline	\noalign{\smallskip}																															
rise1	&13.5   &0		   &2.5$\pm$0.4 	&6.1$\pm$0.3   	&$<0.007$	&0.70$\pm$0.02	 	&8.9$\pm$0.1  &--    &--  &--	   &2.1/80 \\
rise2	&29.3   &0		   &7.0$\pm$2.5	   	&3.5$\pm$0.3    &0.33$\pm$0.04	&1.54$\pm$0.06		 &7.6$\pm$0.2  &--    &--  &--	   &1.5/79 \\
peak	&32.2   &0		   &7.6$\pm$4.0		&3.8$\pm$0.5    &0.8$\pm$0.1	&2.0$\pm$0.09	 	&6.0$\pm$0.3  & --   &--  &--	   &2.5/76 \\
decay1	&24.6   &0		   &3.7$\pm$2.0		&3.1$\pm$0.3    &0.04$\pm$0.02	&0.85$\pm$0.05		 &8.6$\pm$0.1  &--   &--  &--	   &1.2/80 \\
decay2	&15.2   &0		   &2.5$\pm$0.3    	&6.8$\pm$1.1    &0.08$\pm$0.01	&0.86$\pm$0.01	 	&10.2$\pm$0.2  &--    &--  &--	  &2.3/81 \\
decay3	&8.8    &0		   &1.9$\pm$0.9		&8.1$\pm$0.6    &0.04$\pm$0.01	&0.54$\pm$0.01	 	&13.1$\pm$0.3 &--    &--  &--	   &1.3/81 \\
decay4(HB)&1.7  &0		   &7.2$\pm$4.0    	&9.5$\pm$1.1    &0.09$\pm$0.01	&0.21$\pm$0.01   	&20.9$\pm$0.3 &--   &-- &--	   &1.5/85 \\
\hline \hline	\noalign{\smallskip}																															                   
\multicolumn{12}{l}{$*$: including components in Table~\ref{specres2} and \ref{specres3}} \\
\multicolumn{12}{l}{$**$: affected by the mHz QPO. See Fig.~\ref{qpo_0115} and Table~\ref{specres3}. See also \citet{cobu04}}\\
\end{tabular}
\end{table*}
\begin{table}
\caption{Power spectral parameters (centroid frequency, FWHM and $rms$) for the 
$L_{u'}$ component in \exo. Errors are 90\% confidence level.}             
\label{specres2}      
\centering          
\begin{tabular}{lccc}
\hline\hline       	
Spectral	&$\nu_{u'}$ 	&$FWHM_{u'}$ 	 &$rms_{u'}$ \\
region		& (Hz)		& (Hz)	 &(\%)		\\
\hline	\noalign{\smallskip}																						
rise1		&--			&--		&--	\\
rise2		&2.5$\pm$0.1		&5.7$\pm$0.2	&7.2$\pm$0.3	\\
peak		&3.1$\pm$0.6		&6.6$\pm$0.4	&7.1$\pm$1.0   \\
decay1		&2.3$^{+1.1}_{-0.4}$	&6.7$\pm$0.7	&7.8$\pm$2.2	\\
decay2		&--			&--		&--	\\
decay3		&--			&--		&--	\\
decay4		&--			&--		&--	\\
\hline \hline	\noalign{\smallskip}						        															  
\end{tabular}
\end{table}

\begin{table*}
\caption{Power spectral parameters (centroid frequency, FWHM and $rms$) for the 
$L_{mQPO}$, $L_{LF}$ and $L_s$ components. Errors are 90\% confidence level.  The
centroid frequency of the zero-centred Lorentzian was fixed during the fit.
HB stands for horizontal branch.}
\label{specres3}      
\centering          
\begin{tabular}{lc|ccc|ccc|ccc}
\hline\hline       
Spectral&$L_x$	&\multicolumn{3}{c}{$L_{mQPO}$}		&\multicolumn{3}{c}{$L_{LF}$}			&\multicolumn{3}{c}{$L_s$}	\\
region	&$\times 10^{37}$&$\nu_{mQPO}$ 	&$FWHM_{mQPO}$ 	 &$rms_{mQPO}$&$\nu_{LF}$ 	&$FWHM_{LF}$ 	 &$rms_{LF}$	&$\nu_s$ 	&$FWHM_s$	&$rms_s$	\\
	&erg s$^{-1}$ 	& (Hz)		& (Hz)		 &(\%)	& (Hz)		& (Hz)	 &(\%)	&(Hz)		&(Hz)		&(\%)		\\
\hline \hline \noalign{\smallskip}																							
\multicolumn{11}{c}{{\bf \vcas}} \\																							
\hline	\noalign{\smallskip}																						
rise1	 &7.5 	 &--		&-- 		&--		&0    	   	  &0.14$\pm$0.03	&10.7$\pm$0.7	   &--	&-- 	 &--	\\
rise2	 &11.8 	  &--		&-- 		&--		&0.038$\pm$0.008  &0.06$\pm$0.01	&6.3$\pm$0.5	   &--	&--	 &--  \\
peak	 &11.0 	 &3.7$\pm$0.4	&4.3$\pm$0.7  	&8.8$\pm$0.4	&0.006$\pm$0.003  &0.09$\pm$0.01	&8.0$\pm$0.3       &--	&--	 &--   \\   
decay1	 &7.9  	&2.6$\pm$0.4	&5.5$\pm$1.0  	&8.0$\pm$0.8	&0.049$\pm$0.006   &0.06$\pm$0.01	&7.3$\pm$0.6       &--	&--	 &-- \\
decay2	 &6.0  	&--		&-- 		&--		&0.059$\pm$0.005   &0.06$\pm$0.01	&6.3$\pm$0.6	   &--	&--	 &--    \\
decay3 	 &5.0 	&3.6$\pm$0.6	 &4.2$\pm$1.0	&5.4$\pm$0.3	&0.065$\pm$0.003   &0.05$\pm$0.01	&6.8$\pm$0.3       &--	&--	 &-- \\
decay4(HB)&0.9  &--		&--		&--		&0		   &0.03$\pm$0.01	&12.8$\pm$1.0      &--	&--	 &-- \\

\hline	\noalign{\smallskip}																						
\multicolumn{11}{c}{{\bf \bq}} \\																							
\hline	\noalign{\smallskip}																						
rise1	 &13.5  &--		  &--		&--	&0.038$\pm$0.003	&0.059$\pm$0.006 &4.7$\pm$0.1	 &0.22$\pm$0.01    &0.12$\pm$0.01	&4.1$\pm$0.1 \\
rise2	 &29.3  &--		  &--		&--	&0.029$\pm$0.003	&0.079$\pm$0.009 &4.4$\pm$0.1    &0.25$\pm$0.01    &0.26$\pm$0.03	&5.8$\pm$0.3 \\
peak	 &32.2  &--		  &--		&--	&0.010$\pm$0.009	&0.094$\pm$0.020 &4.1$\pm$0.6    &0.23$\pm$0.01    &0.57$\pm$0.03	&8.5$\pm$0.4 \\
decay1	 &24.6  &--		  &--		&--	&0.022$\pm$0.007	&0.078$\pm$0.009 &4.5$\pm$0.4    &0.24$\pm$0.01    &0.15$\pm$0.03	&4.0$\pm$0.3 \\
decay2	 &15.2  &--		  &--		&--	&0.046$\pm$0.002	&0.043$\pm$0.003 &5.3$\pm$0.2    &0.23$\pm$0.01    &0.13$\pm$0.01	&6.2$\pm$0.1 \\
decay3	 &8.8  &--		  &--		&--	&0.045$\pm$0.002		&0.053$\pm$0.005 &9.4$\pm$0.4    &0.23$\pm$0.01    &0.06$\pm$0.02	&3.2$\pm$0.9 \\
decay4(HB)&1.7  &--		  &--		&--	&0.043$\pm$0.005	&0.050$\pm$0.030 &14.4$\pm$0.6	 &--		  &--			&--       \\
\hline \hline	\noalign{\smallskip}						        															  
\end{tabular}
\end{table*}

   \begin{figure*}
   \centering
   \includegraphics[width=17cm]{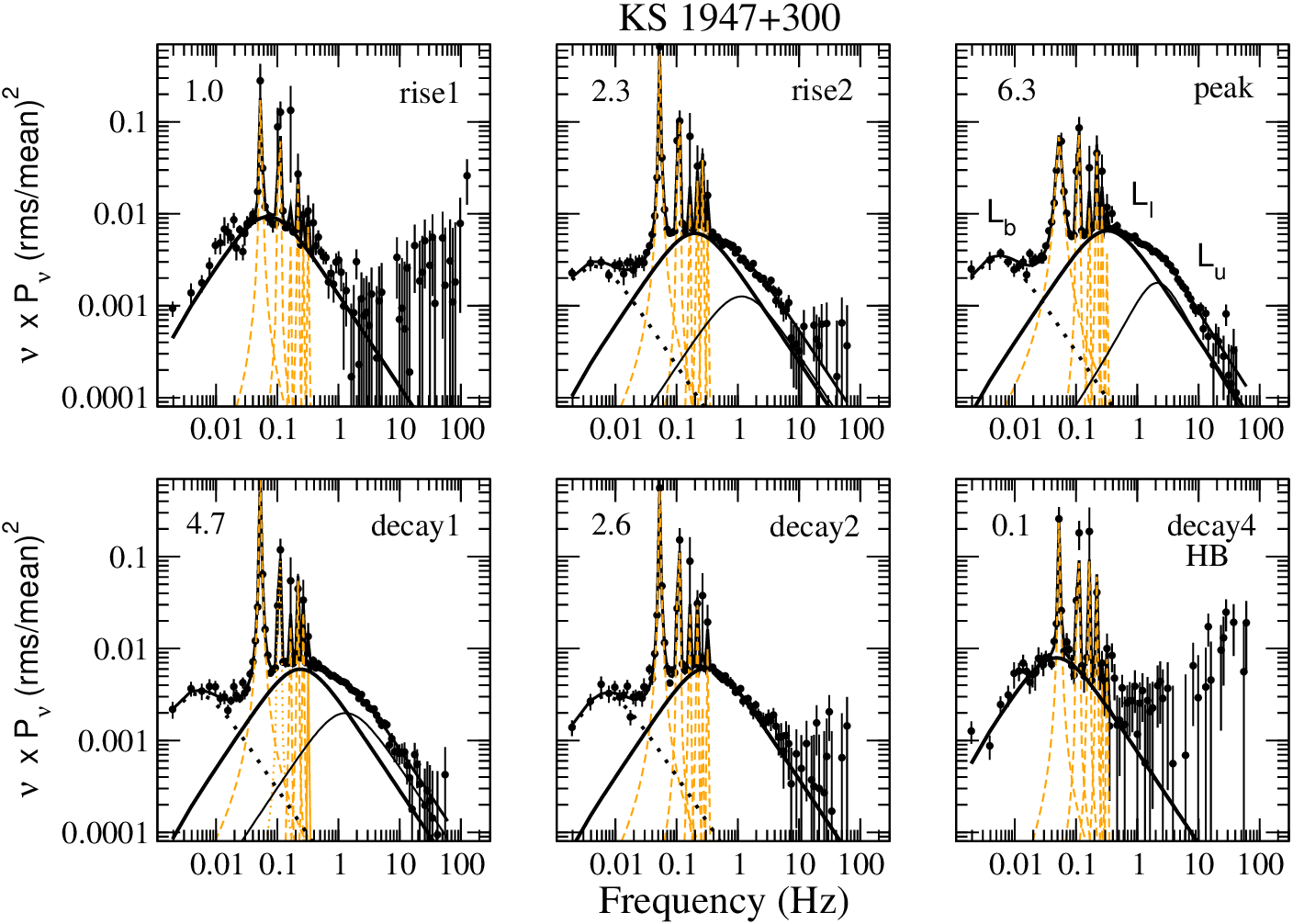} 
      \caption{Power spectra of \ks\ and fit functions of various spectral 
      regions. 
      The number at the top left of each panel is the X-ray luminosity in 
      units of $10^{37}$ erg s$^{-1}$. HB stands for horizontal branch.
      The different lines indicate the individual Lorentzian components of
      the fit: $L_b$ (dotted), $L_l$
      (thick solid black), 
      $L_u$ (thin solid black). The pulse peaks 
      (fundamental and harmonics) are represented by thin dashed orange
      lines.}     
      \label{psd2}
   \end{figure*}
   \begin{figure*}
   \centering
   \includegraphics[width=17cm]{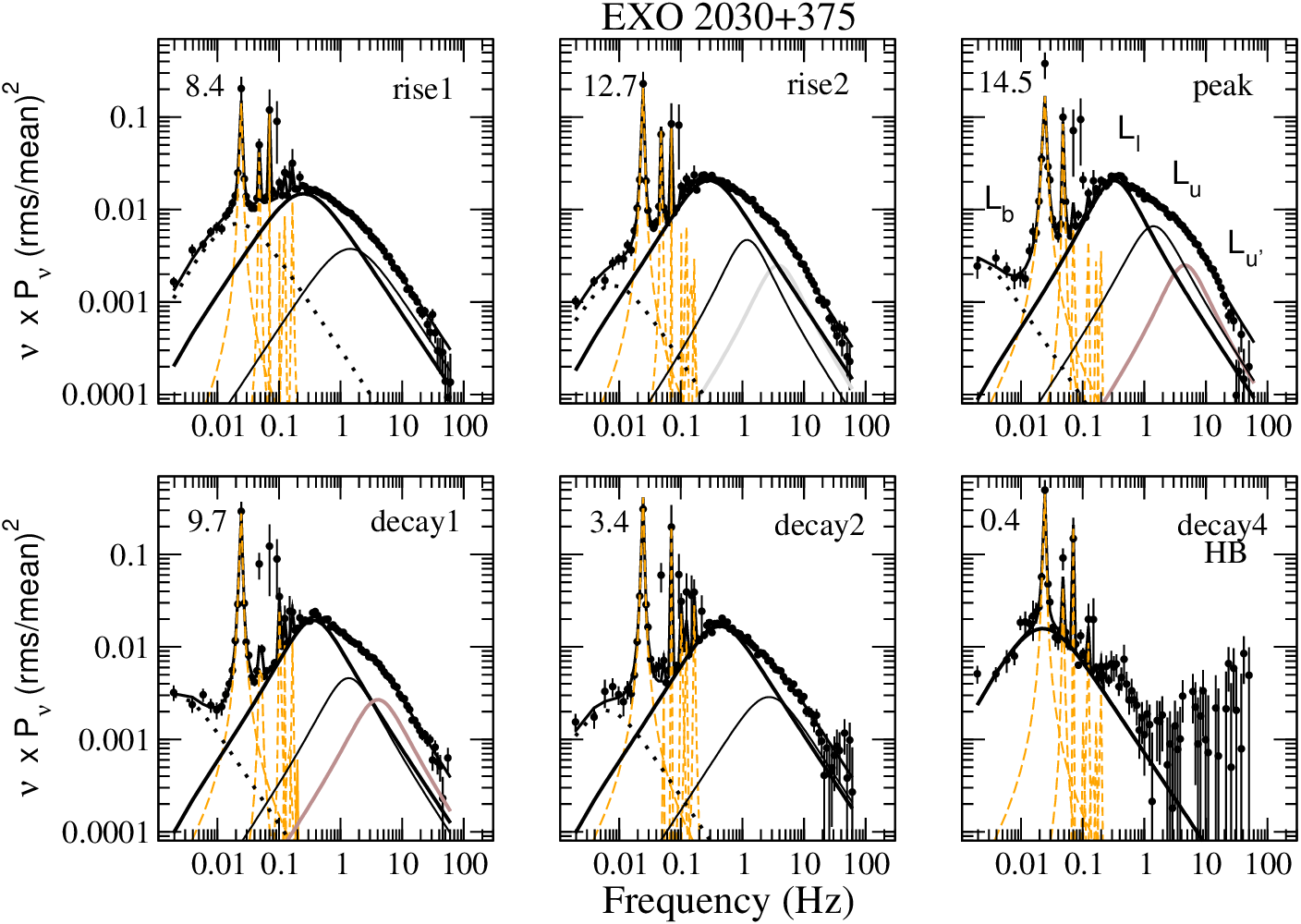} 
      \caption{Power spectra of \exo\ and fit functions of various spectral 
      regions. The number at the top left of each panel is the X-ray luminosity
      in units of $10^{37}$ erg s$^{-1}$. HB stands for horizontal branch.
      The different lines indicate the individual Lorentzian components of
      the fit: $L_b$ (dotted), $L_l$
      (thick solid black), $L_u$ (thin solid black), 
      $L_{u'}$ (thick solid grey). The pulse peaks 
      (fundamental and harmonics) are represented by thin dashed orange
      lines.} 
      \label{psd3}
   \end{figure*}
   \begin{figure*}
   \centering
   \includegraphics[width=17cm]{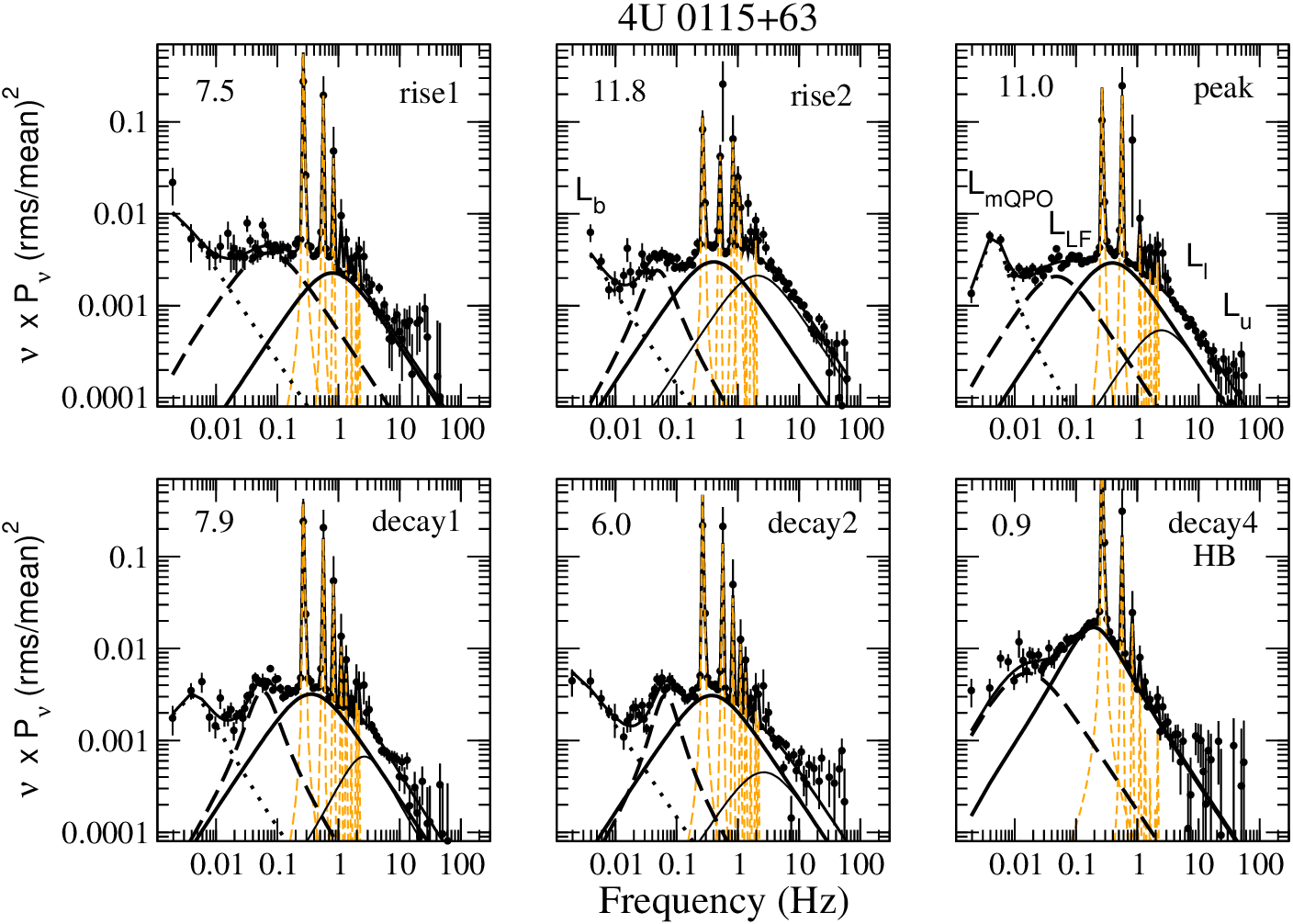} 
      \caption{Power spectra of \vcas\ and fit functions of various spectral 
      regions. The number at the top left of each panel is the X-ray luminosity in 
      units of $10^{37}$ erg s$^{-1}$. HB stands for horizontal branch.
      The different lines indicate the individual Lorentzian components of
      the fit: $L_b$ (dotted), $L_l$
      (thick solid black), $L_{LF}$ (thick dashed), 
      $L_u$ (thin solid black). The pulse peaks 
      (fundamental and harmonics) are represented by thin dashed orange
      lines. In "peak" and "decay1" the $L_b$ component is affected by the
      3 mHz QPO, $L_{\rm VLF}$ (see text and Fig.~\ref{qpo_0115}).} 
      \label{psd1}
   \end{figure*}
   \begin{figure*}
   \centering
   \includegraphics[width=17cm]{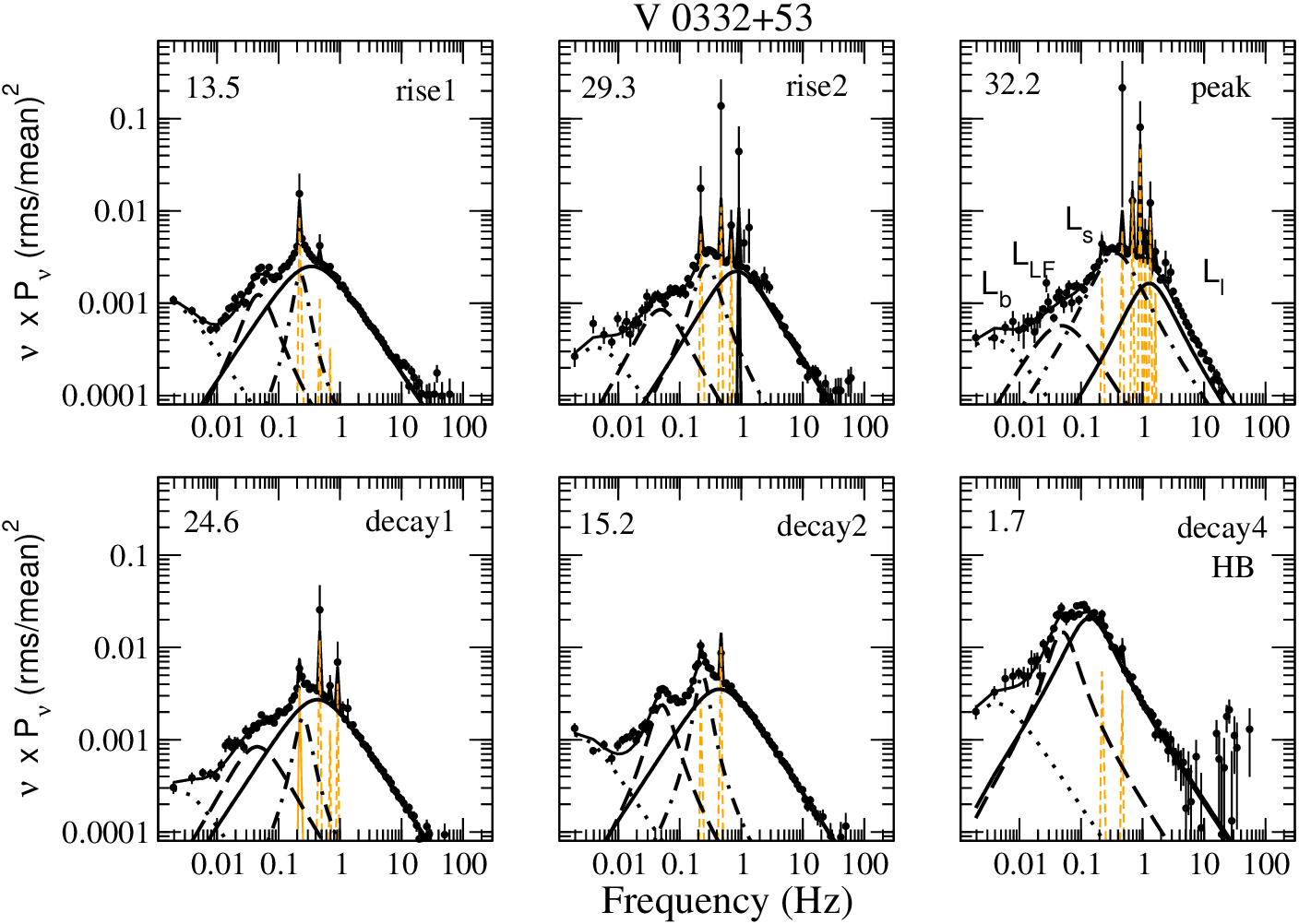} 
      \caption{Power spectra of \bq\ and fit functions of various spectral 
      regions. The number at the top left of each panel is the X-ray luminosity
      in units of $10^{37}$ erg s$^{-1}$. HB stands for horizontal branch.
      The different lines indicate the individual Lorentzian components of
      the fit: $L_b$ (dotted), $L_l$
      (thick solid black), $L_{LF}$ (thick dashed), 
      $L_{s}$ (dashed-dotted), $L_u$ 
      (thin solid black). The pulse peaks (fundamental and harmonics) are 
      represented by thin dashed orange lines.} 
      \label{psd4}
   \end{figure*}

\subsection{Power spectral analysis}
\label{powspec}

Power spectra were created by performing a fast Fourier transform on the
light curves according to the following procedure: one light curve  in
the energy range 2-20 keV (channels 0--49) was extracted for each spectral
region with a time resolution of $2^{-8}$ seconds. Each light curve was
divided into 512-s segments and an FFT was calculated for each segment. The
final power spectra resulted after averaging all the individual power
spectra and rebinning logarithmically in frequency. All power spectra are
presented  in the $rms$ normalisation, where the power integrated over a
certain frequency interval equals the squared $rms$ of the source 
\citep{bell90,miya91}.

The plots of the power spectra are shown in the $\nu \times P_{\nu}$
representation, where each power is multiplied by the corresponding
frequency. This representation helps visualize at which frequency the
contribution to the total $rms$ variability is maximum 
\citep{bell02}. Note that the frequency at which maximum power is attained
in the $\nu \times P_{\nu}$ representation does not equal the centroid
frequency of the Lorentzian, $\nu_0$, but $\nu_{\rm
max}=(\nu_0^2+(FWHM/2)^2)^{1/2}$, where FWHM is the Lorentzian full-width
at half maximum.

The statistical Poisson noise ($P_{N}=2/R_T$, where $R_T$ is the source
count rate) was modified by dead time effects. Dead-time effects are
expected to affect the power spectra at high frequencies, especially during
the high-flux states, where the count rate is over $\sim
1000$ c s$^{-1}$. One way of checking whether the dead-time effects are
important is by measuring the power at high frequencies in a Leahy
normalised power spectrum. The expected power density should be at a level
of 2 \citep{leah83,klis89,jern00}. As an example, the power density of \exo\
during the peak of the outburst ("peak"), calculated for $\nu > 100$ Hz, is
1.935, while that obtained during the end of the outburst ("decay3") is 1.995.
The correction of the Poisson noise $P_N(\nu)$ for dead-time effects was
calculated as described in \citet{nowa99}.

The study of the rapid aperiodic variability in X-ray pulsars is hampered
by the peaks due to the periodic pulsations. The spin period and its
harmonics show up as narrow peaks in the power spectrum that distort the
continuum.  The width of these peaks depends on the frequency resolution,
which in turn, depends on the length of the segments ($\Delta \nu=1/T$,
where $T$ is the total duration of the segment). On one hand, a high
frequency resolution is required in order to have well defined peaks. On
the other hand, higher frequency resolution implies less number of power
spectra to average and worse signal-to-noise, especially at higher
frequencies. As mentioned above we found a good compromise with $T=512$
seconds.  

Multiple Lorentzian profiles provide a good description of the power
spectra in low-mass accreting neutron star binaries  \citep[see
e.g.][]{stra02,stra03,reig04} and black-hole systems \citep[see
e.g.][]{nowa00,bell02,pott03}. To facilitate the comparison between the
noise components and timing rapid variability of all types of X-ray
binaries we have followed this approach here too. The peaks from the
pulsations were fitted to Lorentzian functions with fixed frequency (at the
expected value according to the spin period and its harmonics) and width
(=0.001 Hz $\approx 1/T$).   Tables~\ref{specres1} and ~\ref{specres2} give
the results of the power spectral analysis. Figures \ref{psd1}--\ref{psd4}
show the power spectra and noise components of various spectral regions.
Any given power spectrum can be fitted with the sum of two to four broad
Lorentzian profiles (excluding the peaks), although we identify seven
different types of noise components. 

\subsection{Energy spectra}

The main objective of this work is to investigate the aperiodic
variability of high-mass X-ray pulsars in correlation with the spectral
states as defined in CD/HIDs, in the same way as it has been extensively
done in low-mass X-ray binaries. A detailed spectral analysis is out of
the scope of this work and will be performed in a future paper. The reader
is referred to \citet{wils08} (\exo), \citet{mowl06} (\bq), \citet{naka06}
(\vcas) and \citet{naik06} (\ks) for a detailed analysis of the energy
spectra during type II outbursts. However, for the sake of clarity in the
discussion, it is illustrative to know the X-ray luminosity of the different
spectral regions. Thus we obtained a time-averaged spectrum for each
spectral region. The 3-30 keV spectrum was fitted with an absorbed
power-law model modified at high energies by an exponential cutoff. In
addition, a Gaussian profile at $\sim$ 6.4 keV, that accounts for iron
emission line, was required in most cases. The X-ray flux and luminosity in
the 3-30 keV energy range is given in Table~\ref{reg}. The distance to the
sources is given in Table~\ref{info}. 

\section{Results}
\label{res}

\subsection{Outburst profiles}

The total duration of the \vcas\ outburst was $\sim$55 days ($\sim$ MJD
53245--53300) but the rise was shorter ($\sim$11 days) than the decay
($\sim$27 days). The brightest phase of the outburst displays a multi-peak
profile and lasted for at least 15 days. The source spent about one third
of the time in this bright state. The maximum flux was achieved on
$\sim$MJD53264 and amounted to $\sim$280 mCrab. The PCA observations of
\vcas\ began at the mid point of the rising part of the outburst on
September 6, 2004 (MJD 53254) and finished on October 25, 2004 (MJD
53304). 

The ASM light curves of \ks\ and \exo\ show very similar characteristics,
despite the difference in overall luminosity (the \exo\ outburst was a
factor of 4 brighter). Not only the outburst profile and duration
($\sim$165 days and $\sim$155 days, respectively) are similar but also the
pre- and post-outburst X-ray activity. In these two sources the main
outburst is preceded by a minor one, which started one orbital period
before the onset of the major event. Likewise, the major outburst was
followed by a series of smaller outbursts separated by a time interval
consistent again with the orbital period (40 days in \ks\ and 46 days in
\exo). The outburst profiles are quite symmetric with the rise and decay
lasting for about 60-70 days. The \ks\ ASM light curve contains a gap
during the rise due to its proximity to the Sun at that time.

\bq\ showed the most intense outburst. The maximum flux recorded in the
energy range 1.3-12.1 keV reached $\sim$1.5 Crab. The outburst began in
2004 November (MJD 53330) and reached maximum flux about one month later
($\sim$ MJD 53368). The total duration of the outburst was $\sim 105$ days.
As in \vcas, the decay was slower than the rise. The difference appeared at
the end of the decay, where a longer tail can be seen
(Fig.~\ref{outbprof}).

There appears to be a relationship between the shape and duration of the
outbursts and the orbital period of the system. \ks\ and \exo\ have the
wider orbits and show the longer and more symmetric outbursts. The outburst
profiles of \vcas\ and \bq\ have positive skewed (elongated tail at the
right) and last shorter.  The skewness statistics  (the third moment
about the mean normalised to the standard deviation, 
$\frac{1}{N}\sum{\left(x_i-\bar{x}\right)^3}/\sigma^3$)
gives the following
values for \vcas, \bq, \ks\ and \exo: 0.24, 0.58, 0.16 and --0.035,
respectively. In Be/X-ray binaries, the reservoir of matter available for
accretion onto the neutron star companion comes from the circumstellar disk
around the Be star's equator.  The duration of the X-ray outbursts is
expected to be proportional to the amount of accreted material. Thus longer
outburst duration implies larger disks. This result agrees with the
correlation between the orbital period and strength of the H$\alpha$ line
\citep{reig97,reig07}. The longer the orbital period, the wider the orbit,
the larger the disk. This correlation is explained by the truncation of the
disk by the neutron star, which will be more efficient for narrower orbits.
\citet{okaz01} estimated that the Be discs in \vcas, \bq\ and \exo\ are
truncated at the 4:1 resonance radius, while that of \ks\ at the 3:1
resonance. In terms of stellar radii, assuming the stellar parameters given
in \citet{okaz01}, these resonance radii translate into truncation radii
$\sim$ 4.7 R$_*$, $\sim$ 5.6 R$_*$,  $\sim$ 6.9 R$_*$ and $\sim$ 8 R$_*$
for \vcas, \bq, \exo\ and \ks, respectively. This numbers should be taken
with caution since the calculations are strongly dependent on the disc
viscosity and assume the same disc physical parameters in all systems.

\subsection{Colour analysis}

Be/X-ray binaries show two spectral branches in the CD/HID.  We shall
refer to these branches as the low-intensity or horizontal branch and the
high-intensity or diagonal branch. However, not all sources display the
entire pattern of variability. Sometimes only one branch is visible in the
CD. In \ks, the scatter of the points that populate the low-intensity
branch is too large to clearly define the branch (see Fig.~\ref{cd}). In
\exo, only the hardest (in terms of SC) part of this branch is visible.  

Unless the motion in the horizontal branch is faster than $\sim$1
day$^{-1}$, it seems unlikely that the lack of a complete branch in \ks\
and \exo, at the end of the outburst, is due to observational gaps. At
least in \exo,  the frequency of the PCA observations was $\sim 1$ day$^{-1}$.
More likely, the lack of a clear horizontal branch in the CD of these two
sources can be attributed to the fact that they did not return to
quiescence. Both \ks\ and \exo\ showed smaller outbursts (type I) after the
major outburst (type II) analysed in this work. The X-ray emission level
did not return to zero in these two systems (i.e. below the sensitivity
threshold of the PCA) as can be seen in Fig.~\ref{rate_hr}. The average
count rate of the last observations was 30 c s$^{-1}$ PCU$^{-1}$ in \exo\
and 5 c s$^{-1}$ PCU$^{-1}$ in \ks. In contrast, \vcas\  seems to have gone
under the detectability limit of the instruments. It is reasonable to think
that if the intensity had gone lower, then the horizontal branch in \ks\
and \exo\ would have shown up more clearly. Note that \bq\ also went back
to $\sim$ 0 c s$^{-1}$ and also displays two distinct branches in the CD. 

In LMXBs, the patterns observed in a CD are often also recognizable in the
corresponding HID. Sometimes, however, spectral branches are more distinct
in one of the two diagrams. Whether CD or HID presents the cleanest pattern
depends on source  and on the quality of the data. Figure~\ref{hid}
displays the HID of the four X-ray pulsars analysed in this work. In
contrast to the CD, the two branches are clearly seen for all four sources
when the HC is replaced by the intensity.   The junction of the two
branches occur when the count rate is at $\sim$10\% of the peak value
($\sim$30\% in \vcas).    The second column in Table~\ref{reg} indicates
the branch in which the source lay during those particular observations. 
For practical reasons, the name ``horizontal branch" is also used if
intensity is plotted on a logarithmic scale in the HID, even though in this
case this branch does not appear horizontal.


The low-intensity (horizontal) branch is populated by points with a low
count rate. Since the PCA monitoring began when the outbursts were already
in progress, the low-intensity branch contains points from the end of the
outbursts. The only source for which data prior to the onset of the
outburst are available is \ks. As can be seen in Fig.~\ref{hid}, the data
points (open black circles) that corresponds to the very beginning of the
outburst also lie in the low-intensity branch. Thus we conclude that the
horizontal branch represents the start and end point of the source in
its journey through the CD/HID. In terms of X-ray luminosity, this branch
appears when  $L_X \simless 10^{37}$ erg s$^{-1}$.

The evolution of the X-ray colours as the outburst evolves, particularly
the hard colour (HC), clearly distinguishes \bq\ from the other three
systems.  Unlike, \ks, \vcas\ and \exo, both the SC and HC in \bq\ decrease
as the count rate increases (see Fig.\ref{rate_hr}).  This different
behaviour of the HC implies that the motion in the diagonal branch, as the
count rate increases, is from the bottom right part of the CD to top left
part in \vcas, \ks\ and \exo\ and from bottom left to top right in \bq.
Another important difference is that the low-intensity branch is  harder
than the high-intensity branch in \bq, while it is softer in \exo, \ks\ and
\vcas\ (Fig.~\ref{cd}). In this latter case, the horizontal branch
corresponds to a low/soft (in terms of the hard colour) state that it is
not seen in low-mass X-ray binaries or black-hole systems.

After the peak, the source returns following the reverse track.  The
sources spend most of the outburst in the diagonal branch. The sources stay
in this branch about 70\%--80\% of the duration of the outburst. As can be
inferred from Table~\ref{reg}, typical timescales are  months in the
diagonal branch and weeks in the horizontal branch. 

\vcas\ exhibits hysteresis, that is, the X-ray colours adopt different
values depending on whether the source is on the rise or the decay of the
outburst, despite that the count rate is similar. This effect is also
present in \bq\ and \exo,  although not as significant as in \vcas. It is
absent in \ks\ (see Fig.~\ref{hid}). In \exo\ and \bq\ both X-ray colours
are larger during the rise than during the decay, that is, the sources
present a harder spectrum during the rise. In \vcas, the changes are more
complex. During the rise SC is larger and HC is smaller than during the
decay.  Although the larger error bars in \ks\ might prevent us from
detecting hysteresis in this source, an effect with the amplitude seen in
\vcas\ should also be seen in the SC-intensity diagram of \ks.

Interestingly, cyclotron resonance scattering features have been observed
for all sources except \ks. In particular, \vcas\ shows up to five
cyclotron line harmonics \citep{cobu04}. However, while the energy of the
fundamental line is found at 11 keV in \vcas\ \citep{sant99}, it appears at
27 keV in \bq\ \citep{pott05}. The cyclotron line in \exo, at 11 keV, is
less significant and lacks harmonic content \citep{wils08}.  According to
the definition of the X-ray colours given in Sect.~\ref{col},  the HC in
\vcas\ is  expected to be more strongly affected by the presence of
the cyclotron line and its harmonics, while in \bq\ and \exo\ its
contribution to the variability in HC would be insignificant. Moreover, the
energy of the fundamental cyclotron line in \vcas\ has been seen to
increase from $\sim$11 kev to 16 keV as the X-ray luminosity decreased
below $5\times10^{37}$ erg s$^{-1}$ \citep{naka06}.   
The relationship between hysteresis, as seen in the HID, and cyclotron
lines needs to be verified through detailed spectral analysis.

\subsection{Aperiodic variability}

The power spectra of Be/X-ray binaries is dominated by the  peaks of
the pulse component and its harmonics. In order to study the aperiodic
variability (band-limited noise and possible QPOs), the pulse noise should
be removed. However, this is not an easy task. Some authors substituted the
points contaminated by the pulses and their harmonics by a spline fit to
the remaining points or to a power law connecting the average of a few
points before and after the pulse gap \citep{bell90}. This procedure does
not give satisfactory results if the peaks are too broad. Others simply
remove the points from the power spectra \citep{reig06}. \citet{ange89}
removed the pulse noise after fitting it with a model consisting of the
first 10 harmonics of the coherent pulsar signal and a power law. In this
work we have opted to maintain the peaks and fit them with Lorentzians. In
addition to the distortion introduced by the pulse noise, the power spectra
continuum of Be/X-ray binaries contain significant substructures only
visible in certain states. Sometime the continuum shows bumps and wiggles,
making it difficult to obtain good fits. This difficulty was also found by
\citet{bell90}, who could not fit the power spectra of a number of
high-mass X-ray binaries with simple analytical laws, such as power law and
cutoff, as in low-mass X-ray binaries.

In general, the power spectra of Be/X-ray binaries is characterised
by band-limited noise (BLN), that is, a steeping of the power density
toward high frequency and a flattening toward low frequency.  Although a
double broken power law may give good fits in some cases (\ks) it cannot
fit all sources and all states (\exo). In addition,  all sources present
peaked noise in certain spectral states. 

Despite this complexity, we find that the power spectra of Be/X-ray
binaries can be fitted acceptably with the sum of a small number of
Lorentzian functions.   Ignoring the peaks of the pulse component, 2--4
Lorentzian profiles are needed to obtain acceptable fits. However, we
identify 7 different types of noise components. 4 of these components are
band-limited  (BLN) and 3 QPO. The results of the power spectral fitting
are given in Tables~\ref{specres1}--\ref{specres3}. Figures
\ref{psd1}--\ref{psd4} show the power spectra of the four X-ray pulsars for
various regions. Different line types indicate different noise components
as explained below.

$L_b$ (dotted line) is a zero-centred Lorentzian that describes the noise
below 0.01 Hz. Due to the very low characteristic frequency of this
component (often peaks outside the available frequency range) it is not
always well constrained. In some states of \vcas, $L_{b}$ is modified by
the appearance of a slow QPO.  $L_{b}$ disappears in the horizontal branch
of all sources except in \bq.

$L_{l}$ (thick solid line) is also a broad component with $Q<0.4$ or
centred at zero. Its characteristic frequency shifts to low frequencies
when the count rate is low. This component dominates most of the power
spectrum, covering the range from $\sim$0.1--10 Hz. In some cases (\bq), it
is the only component fitting the high-frequency end of the power spectrum. 

In \vcas, \ks\ and \exo\ one or two more Lorentzians are needed to account
for the high-frequency noise. We shall call these component $L_{u}$ (black
thin solid line) and $L_{u'}$ (grey thick solid line), with characteristic
frequencies $\nu_{u}\approx 1-3$ Hz, $\nu_{u'}\approx 4$ Hz, respectively  
and $Q\approx 0-0.5$. These components only appear when the count rate is
high, i.e., near and at the peak the outburst.

In general, $L_{b}$ accounts for the noise short-ward of the main pulse
peak, while $L_{l}$ and $L_{u}$/$L_{u'}$ fit the power spectrum above the
spin frequency. Together they account for the band-limited noise.

The faster pulsars (\bq\ and \vcas) display peaked noise components, which
we shall refer to as QPO noise, even though the $Q$ value is not always
larger than 2. Table~\ref{specres3} gives the best-fit parameters of these
components. $L_{\rm LF}$ (dashed line in Fig~\ref{psd1} and \ref{psd4}) is
common to the two sources and has a characteristic frequency $\nu_{\rm LF}
\approx 0.05$ Hz and $0.5 \simmore Q_{\rm LF}\simmore 2$. It becomes
stronger as the count rate decreases. Note that although the centroid
frequency of this Lorentzian component increases as the count rate
decreases, its width decreases. The characteristic frequency (i.e. maximum
frequency) remains fairly constant at the value indicated above ($\sim$0.05
Hz). \bq\ shows, in addition,  another QPO (that we shall call $L_{s}$,
dotted-dashed line) whose centroid frequency coincides with the frequency
of the fundamental peak (0.23 Hz) of the pulse period and remains constant
throughout the outburst. The strength of $L_{s}$ increases with count rate,
reaching maximum during the peak of the outburst. However, it disappears in
low-intensity states, when $L_X$ goes below $\sim 10^{37}$ erg s$^{-1}$.
\vcas\ also shows another QPO at $\sim$ 3 mHz. This QPO is even visible in
the light curve (Fig.~\ref{qpo_0115}) and has also been reported in
previous outbursts of the source \citep{hein99,cobu04}. We have called this
component $L_{\rm mQPO}$. 

   \begin{figure}
   \centering
   \includegraphics[width=8cm]{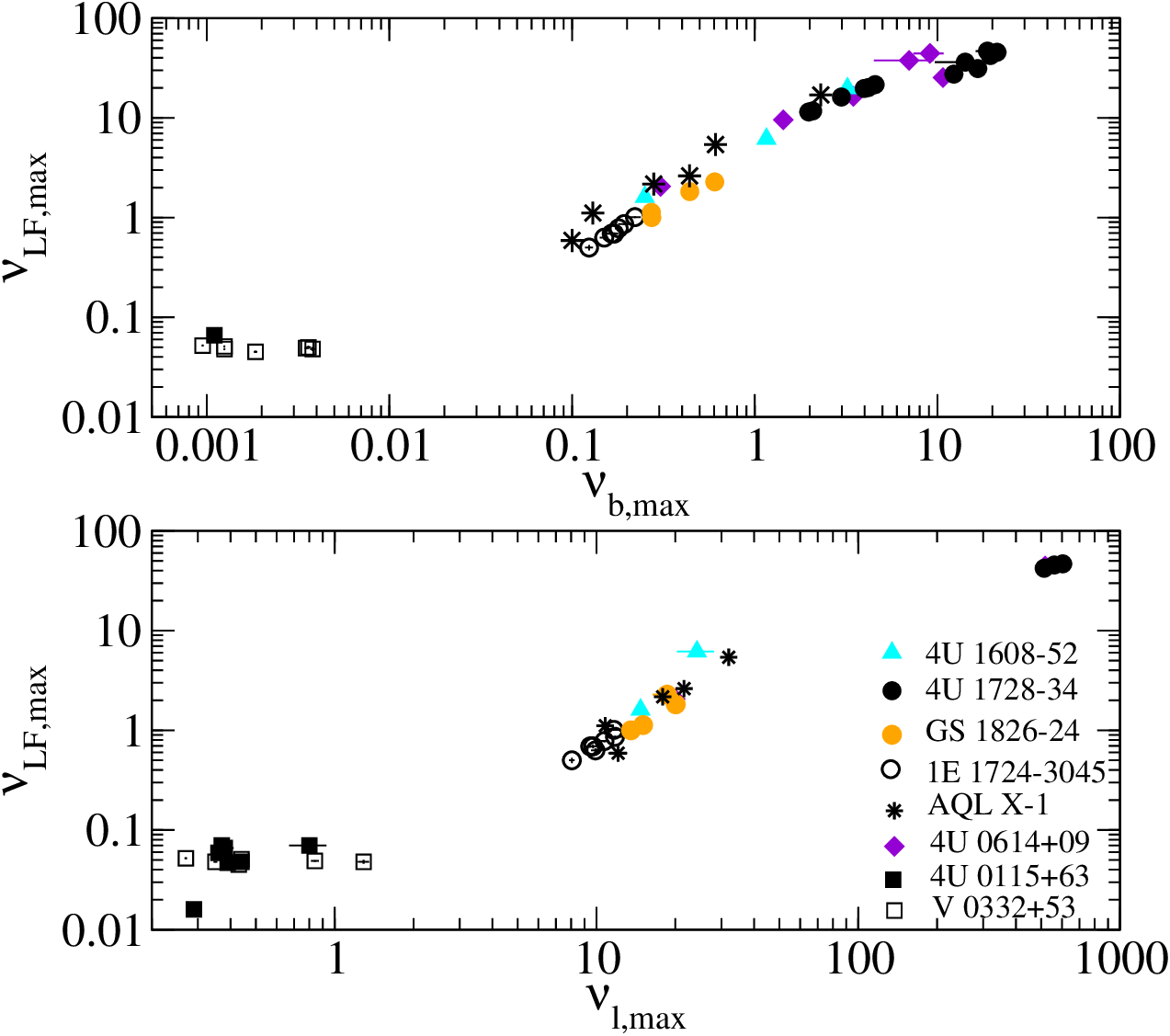} 
      \caption{Correlation between the characteristic frequencies of the
      noise components. Be/X exhibit lower characteristic frequencies than
      LMXBs.
     }     
      \label{freqcor}
   \end{figure}

\section{Discussion}
\label{disc}

In this section, we compare the correlated spectral and timing
variability of high-mass X-ray binaries (HMXBs) and low-mass X-ray
binaries (LMXBs). We comment on the similarities and differences of
these two types of X-ray binaries. 

\subsection{Comparison with low-mass X-ray binaries}
\label{lmxb}

The Be/X-ray binaries that display a wide dynamic range in intensity
(Type II outbursts) trace two-branch patterns in their CD/HID. As in
low-mass systems, the sources do not jump through the diagram but moves
smoothly, following the pattern. The horizontal branch (HB) corresponds to
a low-intensity state and shows the larger fractional $rms$, similar to the
the island state in atolls and horizontal branch in Z sources. This branch
represents the start and end states of the source through the outburst. In
the high-intensity diagonal branch (DB), the noise components display higher
characteristic frequencies and lower $rms$ than in the HB.

The power spectra of LMXBs in the frequency range below 100 Hz are
characterised by \citep{bell02,stra02,stra03,oliv03,klis06}: {\em i)} a
power-law red noise component in the lowest frequency range, traditionally
known as very-low frequency noise (VLFN). This component is weak or absent
in the island and horizontal branches of atoll/Z sources and increases in
strength as the source moves along the branches; {\em ii)} strong
band-limited noise (BLN), consisting of flat-topped noise that breaks at a
frequency $\nu_b \sim 0.01-50$ Hz. This component can be represented by a
zero-centred Lorentzian and fits the low-frequency end of the power
spectrum; {\em iii)} peaked noise at frequency $\sim 5 \nu_b$. When this
peaked noise appears as a QPO then it is called $L_{LF}$ and its
characteristic frequency $\nu_{LF}$. Otherwise, it is normally referred as
$L_h$. Sometimes, both components are present, i.e, a narrow $L_{LF}$ and a
broad $L_h$; {\em iv)} two zero-centred Lorentzians fitting the
high-frequency part of the power spectrum. They normally appear when 
$\nu_b \simless 1$ Hz and are given the names $L_l$ and $L_u$,
respectively. It is always observed that $\nu_b < \nu_{LF} < \nu_l <
\nu_u$.

In Be/X-ray binaries the power spectra can also be well fitted by a
relatively small number of Lorentzian components. Some of these components
can be identified in the power spectra of LMXBs.  As in LMXBs, the
band-limited noise in Be/X-ray binaries is fitted by three broad
Lorentzians: $L_b$ at low frequencies and $L_l$ and $L_u$ at high
frequencies.  In Fig.~\ref{freqcor} we have plotted the relationship
between the characteristic frequencies, i.e. $\nu_{\rm
max}=(\nu_0^2+(FWHM/2)^2)^{1/2}$,  of $L_b$, $L_{LF,h}$ and $L_l$ for the
HMXBs \vcas\ (filled squares) and \bq\ (open squares) and for the LMXBs
\citep{bell02,stra02,stra03,reig04} 4U 1728-34 (black filled circles), 4U
1608-52 (cyan filled triangles), GS 1826-24 (orange filled circles), 1E
1724-3045 (open circles), 4U 0614+09 (diamonds) and Aql X-1 (stars). For
Aql X-1, $L_h$ was plotted. They are known as  the WK relation (after
\citet{wijn99a}, upper panel) and the PBK relation (after \citet{psal99},
lower panel). The fact that the data points corresponding to the Be/X-ray
binaries  fall on the global correlations found for LMXBs, if extrapolated
at low frequencies, supports the identification of the noise components.

There are some important differences in the rapid aperiodic noise
components between HMXBs and LMXBs, namely: {\em i)} the lower
characteristic frequencies, {\em ii)} the lack of a clear correlation of
the noise component parameters with mass accretion rate and {\em iii)} the
noise component, $L_s$, associated with the peaks of the periodic
modulation. 

In accretion-powered X-ray pulsars, QPOs fall in the milliHz range, that is
few order of magnitudes lower than in LMXBs.   There are about 12 high-mass
X-ray pulsars that show QPOs in their power spectra. The frequencies of
these QPOs range from 1--400 mHz \citep{shir02,inam04,mukh06}. Recently,
\citet{kaur07} reported the discovery of a 1.3 Hz QPO in XTE J0111.2--7317,
a Be/X-ray binary located in the SMC. Many of these QPOs have $Q$ values
lower than 2, hence strictly speaking, they should be referred as peaked
noise (conventionally, peaked noise is considered as a QPO when the quality
factor $Q=\nu/FWHM > 2$). However, since these features (also $L_{LF}$ and
$L_s$ in this work) represent a substantial concentration of power in a
limited frequency range they are referred to as QPOs.

If it is assumed that QPOs are formed as a result of Keplerian motion
of inhomogeneities in an accretion disk then the longer implied timescales
of the QPOs in HMXB is somehow expected as the inner radius of the
accretion disk must be larger than the magnetospheric radius ($r_m\sim10^8$
cm). 10-400 mHz
QPOs imply disk radius of the order of $10^8-10^9$ cm  ($R_d\sim
(GM/4\pi^2\nu^2_{QPO})^{1/2}$). The origin of the mHz QPO is, however,
unclear. Neither the Keplerian frequency model \citep{klis87b} nor the beat
frequency model \citep{alpa85} can explain the origin of these QPO in all
systems. QPOs below the neutron star spin frequency cannot be explained by
the Keplerian frequency model because if the Keplerian frequency at the
magnetosphere is less than the spin frequency then the propeller mechanism
would inhibit accretion \citep{stel86}. This is the case of \vcas\ and \bq.
Likewise, the beat frequency model predicts a centrifugal inhibition
threshold that it is at variance with the observations \citep{fing98}. A
third model, the magnetic disk precession model \citep{shir02}, attributes
the mHz QPO in X-ray pulsars to warping/precession modes induced by
magnetic torques near the inner edge of the accretion disk. The
applicability of this model to all sources is not as yet certain
\citep{mukh06}. 

Unlike LMXBs where the power spectral parameters and strength of the
various noise components (anti)correlate with the mass accretion rate,
$\dot{M}$, HMXB noise appears to depend weakly on $\dot{M}$ (we assume that
the accretion rate increases/decreases during the rise/decay of the
outburst), at least at high and intermediate accretion rates. Neither the
characteristic frequencies nor the $rms$ of the Lorentzians that account
for the  band-limited noise show a clear relationship with X-ray flux in
the diagonal branch, despite the change of about 1.5 order of magnitude in
luminosity (see Tables~\ref{specres1}--\ref{specres3}). The lack of such
correlations were also noticed in the 1985 outburst of \exo\
\citep{ange89,bell90}.

$L_s$  is exclusive of accreting X-ray pulsars and it
is associated with the pulse noise. Its central frequency coincides with
the frequency of the fundamental peak of the pulse period and suggests a
strong coupling between the periodic and aperiodic noise components. This
coupling implies that the instabilities in the accretion flow that give
rise to the aperiodic variability must travel all the way down to the
neutron star surface. The reader is referred to \citet{lazz97} and
\citet{burd97} for a detailed study of the coupling between the periodic
and aperiodic variability in X-ray pulsars and to \citet{reig06} for the
particular case of \bq. $L_s$ is absent in the horizontal branch. In the
case of \bq, the disappearance of the pulse peaks in the horizontal branch
is accompanied by the  disappearance of $L_s$, which supports the idea of a
common physical origin.

   \begin{figure}
   \centering
   \includegraphics[width=8cm]{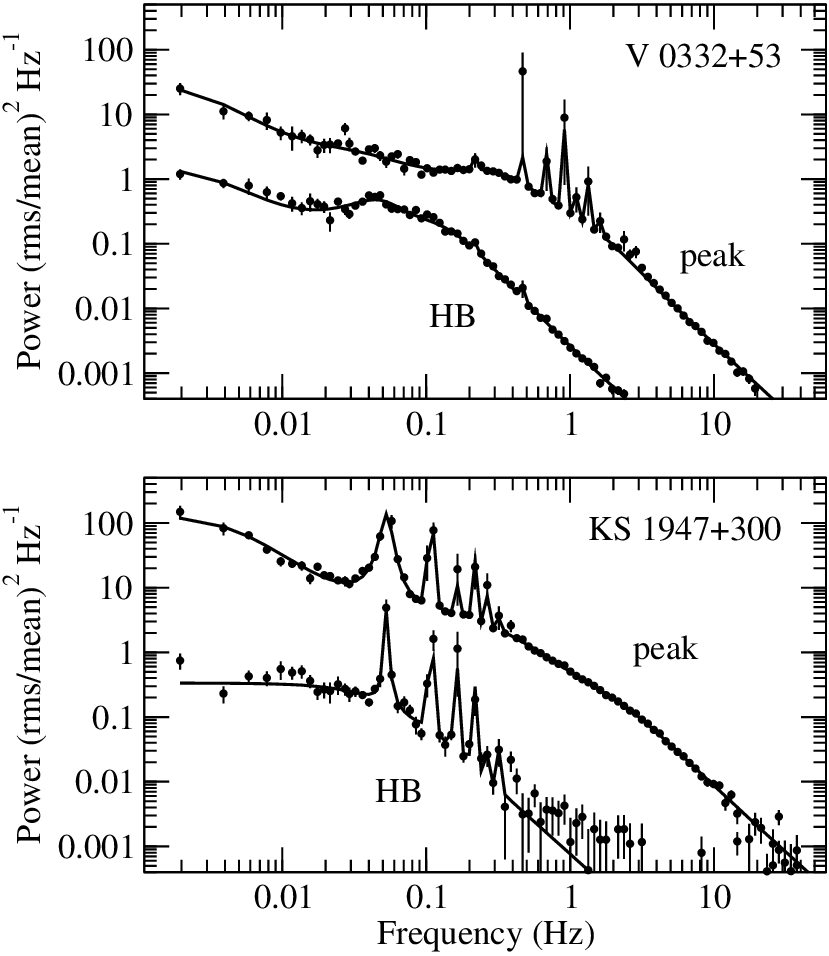} 
      \caption{Representative power spectra of high- and low-intensity
      states. Note the different shape of the noise at low frequencies. 
      Flat-topped noise is seen in the HB of \ks\ (and \exo\ and \vcas) 
      and power-law noise in \bq. Confront this figure with Fig.4 of
      \citet{klis94} and note the similarities between \ks \ and atoll
      sources and between \bq\ and Z sources.
      The "peak" power spectrum was shifted by $\times$100 for easy display.
     }     
      \label{psdexam}
   \end{figure}

\subsection{Are there atoll and Z sources in HMXBs?}

The correlated timing and spectral analysis defines two different types of
Be/X-ray binaries. The differences are not only seen in the shape of the CD
but also in the number and properties of the noise components.

The pattern traced by \bq\ in the CD reminds that of the Z-sources, with
the diagonal branch being the analogue to the normal branch and the HMXB
horizontal  branch branch being the counterpart of the LMXB horizontal
branch. The flaring branch would be missing in \bq. As in Z sources, the
hardest spectrum corresponds to a lower count rate state. Also, the
fractional rms is higher and the characteristic frequencies are lower in
the horizontal branch than in the diagonal branch. 

In \vcas, \ks\ and \exo, the HB also corresponds to a
low-intensity state and also displays the larger $rms$ of the noise
components. However, the HB in these three sources is softer than the
diagonal branch. Likewise, the overall shape and motion along the branches
differs from that seen in \bq\ and LMXBs atoll sources.  The HB in \vcas,
\ks\ and \exo\ corresponds to a low-intensity spectrally soft state not
seen in other types of X-ray binaries. 

Further similarities between \bq\ and Z sources and \vcas, \ks\ and \exo\
and atolls sources are found in the shape of the very-low frequency noise
($\simless 0.1$ Hz). Fig.~\ref{psdexam} shows the power spectra of the HB
and peak of the outburst for \bq\ and \ks. This figure should be compared
to, e.g., Fig.4 in \citet{klis94}. The noise below $\sim$0.1 Hz in the HB
is flat-topped in \ks\ (and in \exo\ and \vcas), as in the island state of
atoll sources, and power-law like in \bq, as in the HB of Z sources.

Initially, the differences between LMXB $Z$ and atoll sources were ascribed
to a higher magnetic-field strength and higher accretion rates in $Z$
sources. However, the detection in atoll sources of QPOs that are
reminiscent of HBO and NBO \citep{ford98,wijn99b} and the existence of sources that
display the two types of behaviour, $Z$ {\em and} atoll
\citep{homa07a,homa07b},
question the validity of the proposed higher magnetic field in the $Z$
sources. Nevertheless, it is worth noting that  \bq\  displays the
most intense outburst and contains the stronger magnetic field of the four
sources. The peak luminosity in \bq\ is about a factor 2, 3 and 5 that of
\exo, \vcas\ and \ks, respectively. The strength of the magnetic field can
be estimated from the energy of cyclotron resonant scattering features
(CRSF) through the relation 

\[E_{\rm cyc}=11.6\frac{B}{10^{12} {\rm G}} (1+z)^{-1} \,\, {\rm keV}\]

\noindent where $B$ is the magnetic field and $z$ is the gravitational
redshift. As mentioned above, \ks\ is the only source for which no CRSF has
been reported, while for \vcas\ $E_{\rm cyc}=12$ keV \citep{sant99}, for
\exo\ $E_{\rm cyc}=11$ keV \citep{wils08}, and for \bq\ $E_{\rm cyc}=27$
keV \citep{pott05}. Thus the magnetic field in \bq\ is about 2--3 times
stronger than in the other three systems.

It is tempting to attribute the differences between Be/X-ray binaries to
the source brightness and magnetic field strength. The power-law noise at
low frequencies in the HB of \bq\ could be due to the larger intensity and
the presence of $L_S$ noise to a higher magnetic field.  \vcas, \ks\ and
\exo\ show flat-topped noise and lack $L_s$ component. However, with only
one source of this type no definitive conclusion can be drawn about whether
the source brightness and magnetic field strength allows the distinction
between different types of systems among HMXB. Future observations will
clarify this issue.

\section{Conclusions}
\label{con}

We have performed, for the first time, a systematic analysis of the spectral
(through the study of hardness ratios) and rapid aperiodic variability of
four Be/X-ray pulsars during type II outbursts. Our aim was to characterise the
rapid aperiodic variability in correlation with the position in the
colour-colour diagram as it has been done for black-hole systems and
low-mass X-ray binaries. 

The correlated spectral-timing behaviour in Be/X-ray binaries shares a
number of similarities with low-mass X-ray binaries, namely, {\em i)}
existence of spectral branches in the CD/HID. At high and intermediate flux
the sources move along the diagonal branch; at very low count rate the soft
colour decreases, while the hard colour remains fairly constant defining a
horizontal branch {\em ii)} smooth motion, i.e. without jumps, in the
CD/HID, {\em iii)} the low-intensity states are more variable (in terms of
fractional rms), {\em iv)} description of power spectra in terms of a small
number of Lorentzian components, {\em v)} similar BLN namely, broad
Lorentzians $Q \sim 0-0.3$ describing the low-frequency and high-frequency
noise ($L_b$, $L_l$, $L_u$) and peaked noise in between ($L_{LF}$) and {\em
vi)} flat-topped noise at the lower frequencies in the horizontal branch
that turns into power-law noise in the diagonal branch.

There are also important differences: {\em i)} different patterns in the
CD/HID. Unlike low-mass X-ray binary and black hole systems, the
low-intensity state is associated with a soft state (except for \bq), {\em
ii)} slower motion along the spectral branches, hours to days in LMXBs,
weeks to months in HMXBs, {\em iii)} the characteristic time scales implied
by the noise components are about one order of magnitude longer in HMXBs
(e.g. mHz QPOs), {\em iv)} $L_S$ noise in HMXBs and {\em v)} no apparent
correlation between the power spectral parameters (characteristic
frequencies, $rms$) and mass accretion rate.

Although we have analysed all Be/X-ray binaries that showed type II
outbursts in the period 1996--2007 and had good data coverage and
sampling, the number of sources is not large enough to draw definite
conclusions on the underlying physical parameters that may explain the
differences between the two subgroups. Increasing the number statistics is,
however, difficult given the transient nature of their X-ray emission and
unpredictability of the type II outbursts.   

Even though all four sources analysed in this work are X-ray pulsars, it is
not clear whether the high-mass X-ray binaries with supergiant companions
show a similar behaviour. In these systems accretion occurs via the strong
stellar wind of the massive companion, which contrasts with accretion via a
disc, as it is believed to occur in Be/X-ray binaries during outburst.
The problem of this type of studies in supergiant X-ray binaries is that
they are rather stable in long time scales and do not show a wide range in
X-ray flux.

\begin{acknowledgements}

This research was supported by the European Union Marie Curie grant
MTKD-CT-2006-039965. This research has made use of NASA's Astrophysics Data
System Bibliographic Services and of the SIMBAD database, operated at the
CDS, Strasbourg, France. The ASM light curve was obtained from the
quick-look results provided by the ASM/RXTE team.

\end{acknowledgements}

\end{document}